\documentclass[]{jfm}
\usepackage{color}
\usepackage[dvipsnames]{xcolor}
\usepackage{tikz}
\usetikzlibrary{shapes}
\usepackage{xcolor,fontawesome}
\usepackage{graphicx}
\usepackage{ccaption}
\usepackage{newtxtext}
\usepackage{newtxmath}
\usepackage{natbib}
\usepackage{hyperref}
\usetikzlibrary[plotmarks] 
\usepackage{subcaption}
\usepackage{siunitx}

\hypersetup{
    colorlinks = true,
    urlcolor   = blue,
    citecolor  = black,
}

\newcommand{\RomanNumeralCaps}[1]
\linenumbers

\usepackage{amssymb}
\usepackage{pifont}

\definecolor{tabblue}{rgb}{0.12156863, 0.46666667, 0.70588235}

\definecolor{taborange}{rgb}{1.  , 0.49803922, 0.05490196}

\definecolor{tabgreen}{rgb}{0.17254902, 0.62745098, 0.17254902}

\definecolor{tabred}{rgb}{0.83921569, 0.15294118, 0.15686275}

\definecolor{tabpurple}{rgb}{0.58039216, 0.40392157, 0.74117647}

\definecolor{tabpink}{rgb}{0.89019608, 0.46666667, 0.76078431}

\definecolor{tabcyan}{rgb}{0.09019608, 0.74509804, 0.81176471}

\definecolor{tabbrown}{rgb}{0.54901961, 0.3372549 , 0.29411765}

\definecolor{tabolive}{HTML}{bcbd22}

\newcommand{\symbolneight}[1][1]{%
  \raisebox{0.5pt}{\tikz{\node[draw,scale=0.5,regular polygon, regular polygon sides=4,draw opacity=0, fill=tabblue, fill opacity= #1](){};}}%
}

\newcommand{\symbolnfour}[1][1]{%
  \raisebox{0pt}{\tikz{\node[draw,scale=0.4,regular polygon, regular polygon sides=3,draw opacity=0, fill=taborange,rotate=180, fill opacity= #1](){};}}%
}

\newcommand{\symbolzero}[1][1]{%
  \raisebox{0.5pt}{\tikz{\node[draw,scale=0.5,diamond,draw opacity=0,fill=tabgreen, fill opacity= #1](){};}}%
}

\newcommand{\symbolfour}[1][1]{%
  \raisebox{0.5pt}{\tikz{\node[draw,scale=0.4,regular polygon, regular polygon sides=3, draw opacity=0,fill=tabred, fill opacity= #1](){};}}%
}

\newcommand{\symboleight}[1][1]{%
  \raisebox{0pt}{\tikz{\node[draw,scale=0.5,regular polygon, regular polygon sides=6,draw opacity=0, fill=tabpurple, fill opacity= #1](){};}}%
}

\newcommand{\symbolzpg}[1][1]{%
  \raisebox{0pt}{\tikz{\node[draw,scale=0.5,circle,draw opacity=0, fill=black, fill opacity= #1](){};}}%
}

\newcommand{\symbolzpgsw}[1][1]{%
  \raisebox{0pt}{\tikz{\node[draw,scale=0.5,circle,draw=black, draw opacity= #1](){};}}%
}

\newcommand{\symbolnfourlower}[1][1]{%
  \raisebox{0pt}{\tikz{\node[draw,scale=0.4,regular polygon, regular polygon sides=3,draw opacity=0, fill=tabpink,rotate=90, fill opacity= #1](){};}}%
}

\newcommand{\symbolneightlower}[1][1]{%
  \raisebox{0pt}{\tikz[every node/.style={kite, draw}]{\node[draw,kite vertex angles=60, draw opacity=0, fill=tabcyan, scale=0.4, fill opacity= #1](){};}}%
}

\newcommand{\symbolntenlower}[1][1]{%
  \raisebox{0pt}{\tikz{\node[draw,scale=0.4,regular polygon, regular polygon sides=3,draw opacity=0, fill=tabbrown,rotate=270, fill opacity= #1](){};}}%
}

\newcommand{\symbolneightsw}[1][1]{%
  \raisebox{0.5pt}{\tikz{\node[draw,scale=0.5,regular polygon, regular polygon sides=4,draw=tabblue, draw opacity= #1](){};}}%
}

\newcommand{\symbolnfoursw}[1][1]{%
  \raisebox{0pt}{\tikz{\node[draw,scale=0.4,regular polygon, regular polygon sides=3,draw=taborange,rotate=180, draw opacity= #1](){};}}%
}

\newcommand{\symbolzerosw}[1][1]{%
  \raisebox{0.5pt}{\tikz{\node[draw,scale=0.5,diamond,draw=tabgreen, draw opacity= #1](){};}}%
}

\newcommand{\symbolfoursw}[1][1]{%
  \raisebox{0.5pt}{\tikz{\node[draw,scale=0.4,regular polygon, regular polygon sides=3, draw=tabred, draw opacity= #1](){};}}%
}

\newcommand{\symboleightsw}[1][1]{%
  \raisebox{0pt}{\tikz{\node[draw,scale=0.5,regular polygon, regular polygon sides=6,draw=tabpurple, draw opacity= #1](){};}}%
}

\newcommand{\symbolneightlowersw}[1][1]{%
  \raisebox{0pt}{\tikz[every node/.style={kite, draw}]{\node[draw,kite vertex angles=60, draw=tabcyan, scale=0.4, draw opacity= #1](){};}}%
}

\newcommand{\symbolntenlowersw}[1][1]{%
  \raisebox{0pt}{\tikz{\node[draw,scale=0.4,regular polygon, regular polygon sides=3, draw=tabbrown,rotate=270, draw opacity= #1](){};}}%
}

\title{Effects of pressure gradient histories on skin friction and mean flow of high Reynolds number turbulent boundary layers over smooth and rough walls}

\author{T. Preskett\aff{1}
  \corresp{\email{tdp1g17@soton.ac.uk}}, M. Virgilio\aff{1}, P. Jaiswal\aff{1}, B. Ganapathisubramani\aff{1}}

\affiliation{\aff{1}Aerodynamics and Flight Mechanics Research Group, \\ University of Southampton,
Southampton SO17 1BJ, UK}

\begin{document}

\maketitle

\begin{abstract}
Experiments are conducted over smooth and rough walls to explore the influence of pressure gradient histories on skin friction and mean flow of turbulent boundary layers. Different pressure gradient histories are imposed on the boundary layer through an aerofoil mounted in the freestream. Hot-wire measurements are taken at different freestream velocities downstream of the aerofoil where the flow has locally recovered to zero pressure gradient but retains the history effects. Direct skin friction measurements are also made using oil film interferometry for smooth walls and a floating element drag balance for rough walls. The friction Reynolds number, $Re_\tau$, varies between $3000$ and $27000$, depending both on the surface conditions and the freestream velocity ensuring sufficient scale separation. Results align with previous findings, showing that adverse pressure gradients just upstream of the measurement location increase wake strength and reduce the local skin friction while favourable pressure gradients suppress the wake and increase skin friction. The roughness length scale, $y_0$, remains constant across different pressure gradient histories for rough wall boundary layers. Inspired by previous works, a new correlation is proposed to infer skin friction based on the mean flow. The difference in skin friction between an arbitrary pressure gradient history and zero pressure gradient condition can be predicted using only the local wake strength parameter ($\Pi$), and the variations in wake strength for different histories are related to a weighted integral of the pressure gradient history normalised by local quantities. This allows us to develop a general correlation that can be used to infer skin friction for turbulent boundary layers experiencing arbitrary pressure-gradient histories.

\end{abstract}


\section{Introduction}\label{sec:introduction}

Turbulent boundary layers (TBL) at high Reynolds numbers over smooth and rough walls are common in a variety of engineering applications and natural environments. The variation of the mean velocity ($U$) with wall-normal position ($y$) of a turbulent boundary layer with thickness ($\delta$) in the outer region of smooth/rough walls is usually represented with a log-wake composite profile, 

\begin{eqnarray}
 U^+_{smooth} & = & \frac{1}{\kappa} ln(y^+) + B + \frac{\Pi}{\kappa}W(\eta), ~{\rm with}, \eta = \frac{y}{\delta}     \label{eq:complete_velocity_profile_SW} \\
 U^+_{rough} & = & \frac{1}{\kappa} ln\left(\frac{y-d}{y_0}\right) + \frac{\Pi}{\kappa} W(\eta) ~{\rm with}, \eta = \frac{y-d}{\delta-d}  \label{eq:complete_velocity_profile_RW} \\
{\rm where,}~W(\eta) &=& 2 \eta^2 (3- 2 \eta) - \frac{1}{\pi} \eta^2 (1- \eta)(1 - 2 \eta) \label{eq:polynomial_wake}
\end{eqnarray}

\noindent
where $U^+ = U/U_\tau$ is the mean velocity scaled with skin friction velocity ($U_\tau = \sqrt{\tau_w/\rho}$, $\tau_w$ is the wall-shear-stress and $\rho$ is the fluid density), $y^+ = yU_\tau/\nu$ is the inner scaled wall-normal position ($\nu$ is the kinematic viscosity), $\kappa$ and $B$ are von K\'{a}rm\'{a}n constant and smooth wall intercept, typically set at 0.39 and 4.3 respectively (\citealt{marusic_2013}). For a rough wall, $y_0$ is the roughness length, and $d$ is the zero-plane displacement (or a virtual origin for the log region), and both are flow/surface-specific quantities. The outer region for a rough wall starts typically $3-5$ representative roughness heights above the surface (\citealt{schultz_2007,jimenez_2004,chung_2021}). The roughness length scale ($y_0$) is analogous to Nikuradse's equivalent sandgrain roughness ($k_s$) and are trivially related to each other (\citealt{chung_2021}). Finally, $\Pi$ is Cole's wake strength and $W$ is the functional form for the wake. There are several implementations for the wake function ($W$), and in this work, we only consider the form in equation \ref{eq:polynomial_wake} provided by \cite{lewkowicz_1982}.

\citet{townsend_1956} first proposed outer-layer similarity between smooth and rough walls (at least for zero-pressure-gradient - ZPG flows) where the flow in the outer region is not different between different surface conditions and the influence of roughness is limited to the near wall region (which is below the outer region). This implies that turbulent motions (mean flow, turbulence statistics and even structures) may behave similarly regardless of surface conditions in the region outside of the immediate roughness layer, at sufficiently high Reynolds numbers (\citealt{schultz_2007}). For the mean flow, this implies that the outer-wake region (i.e. the value of $\Pi$ as well as the function $W$) is similar between smooth and rough walls. This is typically assessed with the mean velocity in deficit form as given by equation \ref{eq:deficit_profile}, 

\begin{equation}
    \frac{U_{99} - U}{U_\tau} = -\frac{1}{\kappa} ln\left(\frac{y-d}{\delta - d}\right) + \frac{\Pi}{\kappa} \left[2 - W\left(\frac{y-d}{\delta-d}\right)\right] 
    \label{eq:deficit_profile}
\end{equation}

where $U_{99}$ is the boundary layer edge velocity. 

Previous works have shown that mean profiles in deficit form do indeed collapse between smooth and rough walls for ZPG flows provided the representative roughness height ($k$) is small compared to boundary layer thickness ($\delta$). Different studies have reported different thresholds for this ratio ranging from $k/\delta$ = 0.02 to 0.1 (\citealt{jimenez_2004,castro_2007}). This threshold appears to depend on the type of roughness as well as the scale separation achieved in the flow (i.e. Reynolds number). The presence of outer-layer similarity together with the knowledge of $y_0$ (roughness length of a given surface)allows us to develop models that can be used to calculate skin friction and other boundary layer parameters at higher Reynolds numbers (\citealt{castro_2007,monty_2016}). 

Most realistic systems do not operate under ZPG conditions due to surface curvature or external flow effects. When studying pressure gradient (PG) flows, it is essential to distinguish between equilibrium and non-equilibrium conditions since that could affect the nature of outer-layer similarity as well as other characteristics of the boundary layer flow. Equilibrium flows are those in which the mean velocity profiles and flow statistics are invariant with the streamwise position. However, the only true equilibrium flow is that of a smooth wall sink flow (\citealt{townsend_1956,rotta_1962}). A near equilibrium boundary layer is defined by \citet{marusic_2010} as one where the mean velocity deficit exhibits self-similarity at a high enough Reynolds number. For a near equilibrium flow, the Clauser parameter $\beta$, defined as in equation \ref{eq:beta_rc}, must be constant \citep{clauser_1954}. 
\begin{equation}
    \beta = \frac{\delta^*}{\tau_w}\frac{dP}{dx}
    \label{eq:beta_rc}
\end{equation}

where $\beta$ is the pressure gradient parameter, $\delta^*$ is the displacement thickness, $\tau_w$ is the wall shear stress and $dP/dx$ is the streamwise pressure gradient. It remains unclear how non-equilibrium conditions, i.e. streamwise variation in $\beta$ generated by pressure gradients, affect both the flow over a rough surface and the wall similarity hypothesis. 

Extensive studies have been performed on smooth wall flows under various pressure gradients. However, work on rough walls (with comparable pressure gradients) is more limited. Adverse pressure gradients (APG) have received more attention due to their association with flow separation and the resulting increase in drag. The most prominent effect of an APG is its impact on the wake region. Hot-wire measurements of \cite{monty_2011} and laser doppler velocimetry (LDV) measurements in the work of \cite{aubertine_2005} over smooth walls show larger wake strengths (i.e. larger values of $\Pi$) with APG. Smooth wall DNS results provided by \cite{lee_2009} and validated by  \cite{monty_2011} showed that the wall-normal extent of the log region is limited under APG conditions. The skin friction coefficient has been found to decrease with APG in the works of \cite{shin_2015_apg} and \cite{volino_2020}; however, it is difficult to quantify these effects as these results were derived from the mean velocity profile. The work of \cite{monty_2011} also showed this decrease using independent skin friction measurements but only for a smooth wall case. 

The earliest experiments on rough wall boundary layers with APG were carried out by \citet{perry_1963}, who reported that the roughness function was independent of the pressure gradient the flow experienced. However, this conclusion has been challenged by more recent works. The experiments of \citet{pailhas_2008} found that an APG affected the value of $k_s$ (or $y_0$). Particle image velocimetry (PIV) of flows over ribs, carried out by \citet{tsikata_2013}, concluded that the combined effect increases $k_s$, while amplifying the wake and reducing the length of the log region. \cite{tay_2009_apg} found that $k_s$ also increases with APG and that the effects on a TBL of both roughness and APG augment each other. Likewise, \citet{tachie_2007} concluded that the combined effect of roughness and APG was greater than that of roughness on its own, resulting in a larger roughness sublayer. Hot-wire measurements by \citet{shin_2015_apg} concluded differently that APGs reduce the effect of rough walls and reduce the skin friction compared to zero pressure gradient. \citet{song_2002} by means of Laser doppler anemometry (LDA) showed an earlier separation on a rough wall with APG compared to a smooth wall, supported by the work of \citet{aubertine_2004}. Turbulent boundary layers with favourable pressure gradients (FPG) have been studied with less attention in the literature. Over smooth walls, FPG boundary layers have been shown to increase the log layer length due to relaminarisation effects induced by the flow acceleration (\citealt{piomelli_2000}). As one may expect, the DNS simulations of \cite{yuan_2015} showed that TBLs under FPGs do not relaminarise in the case of rough wall flows. The work of \cite{tay_2009_fpg} and \cite{ghanadi_2022} showed FPGs also result in thinner boundary layers with a smaller wake strength (i.e. smaller $\Pi$), if compared to a ZPG case. Skin friction has been shown to increase over smooth and rough walls by the works of \cite{tay_2009_fpg} and \cite{shin_2015_fpg}.

The aforementioned studies have investigated the effects of pressure gradients over rough walls, however, often for a single pressure gradient type. Very few studies have examined the combined effect of FPGs and APGs. Limited work has been carried out on this subject over smooth walls. The work of \cite{bobke_2017} looks at different non-equilibrium APG histories over a flat plate and an aerofoil surface. They found that different streamwise developments of $\beta$ imply that even for equal flow states ($Re_{\tau}$ and $\beta$) at the measurement location result in different velocity profiles and turbulence statistics. This phenomenon is due to the historical effects of the pressure gradients on TBL development, which is also shown by \cite{vila_2017}. The effects of pressure gradient histories on flows over rough walls are poorly understood, yet it is crucial for predicting drag over rough surfaces and enhancing system efficiencies. The work of \citet{fritsch_2022} and \citet{vishwanathan_2023_IJHF} as well as \citet{volino_2023} demonstrated the variation in $k_s$ under different pressure gradients. These studies had contrasting conclusions with one suggesting that $k_s$ (or $y_0$) is independent of pressure gradient histories (although the values of $\beta$ explored were not very strong and the variance in $k_s$ was much as 50\% across cases, but, without any specific trends with pressure gradient) while the other showed that $k_s$ increases with FPG and decreases with APG. \cite{vishwanathan_2023_IJHF} suggested that the variation in $k_s$ is due to the choice of extent of log region during the fitting process, which is necessary to determine $k_s$ (or $\Delta U^+$). \cite{volino_2023} also indicated a dependence of $k_s$ on $k/\delta$, suggesting that these results may reflect a lack of scale separation. In should be noted that all these studies have significant uncertainty in their results also due to indirect wall shear stress measurements. Therefore, any fitting process and determination of parameters depend on the value of skin friction.  

There is still a need for high-fidelity experimental data of boundary layers over smooth and rough walls experiencing different pressure gradient histories. This type of data can then be used for developing new predictive models for skin friction where the non-equilibrium effects can be captured. With access to high-fidelity data with sufficient scale separation, it may be possible to develop these models following the approaches of \cite{perry_2002} or \cite{castro_2007},  where bulk boundary layer characteristics can be determined using momentum integral approaches. Alternately, an empirical relationship for skin friction can be developed (for example, \citealt{vinuesa_2017}). Overall, the above review points to several open questions that need to further explored. These include, $(i)$ Does the value of $y_0$ (or $k_s$) depend on scale separation or pressure gradient history? $(ii)$ Can we translate information from smooth wall flows with a given pressure gradient history to rough walls with similar histories?, $(iii)$ Is it possible develop new prediction/correlation models that can  infer boundary layer properties with limited measurements? and $(iv)$ Would the history or strength of pressure gradients applied to smooth/rough walls influence the applicability of these models?

In this study, we address some of these questions through detailed measurements of smooth and rough wall boundary layers in a region where the flow locally has zero pressure gradient, but, has experienced very different pressure gradient histories.  Hot-wire, oil film interferometry and floating-element drag-balance measurements are carried out to gain new insights on the mean flow. Based on the observations from the data, we develop a correlation model that can be used to predict the local skin friction coefficient that includes history effects for smooth and rough walls. The paper is organised in the following sections. Section \ref{sec:Methodology} discusses the experimental methods used and we present the mean velocity and skin friction data in section \ref{sec:Results}. The data is further reduced and analysed in the context of predictive models in section \ref{sec:Models} with final conclusions and further recommendations in section \ref{sec:Conclusions}.

\section{Methodology}\label{sec:Methodology}
This section describes the experiments carried out and the pressure gradient histories imposed on smooth and rough wall turbulent boundary layers. 

\subsection{Facility}\label{sec:Facility}

\begin{figure}
    \centering
    \includegraphics[width = \textwidth]{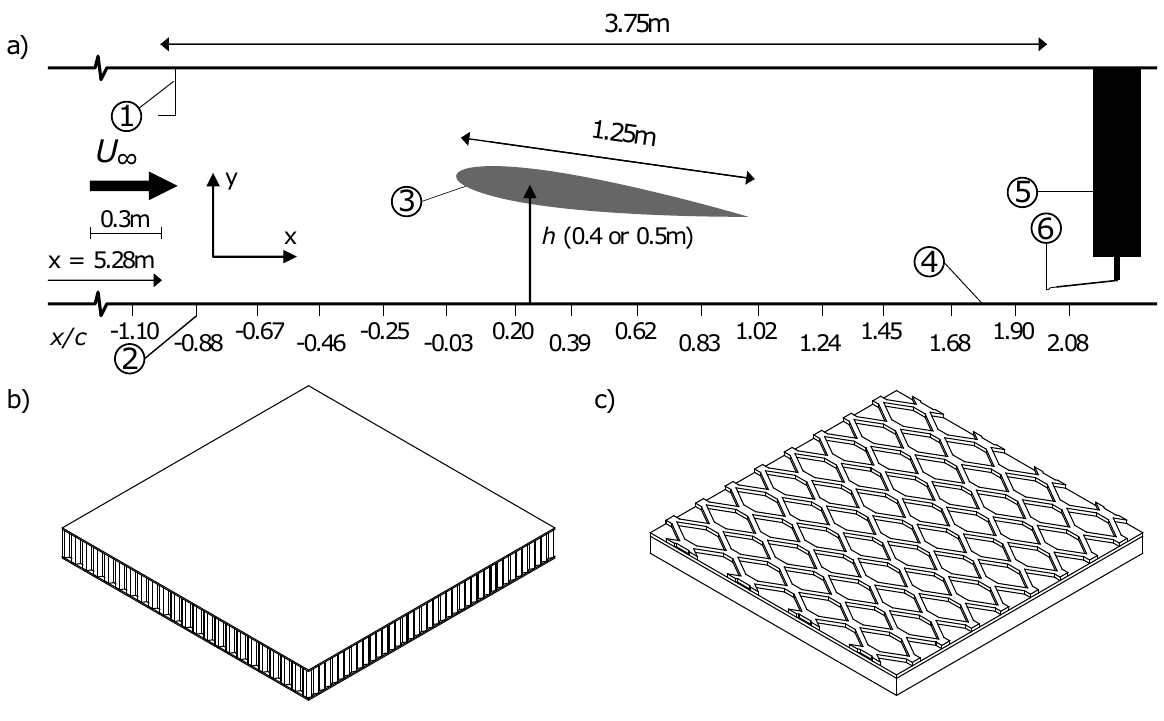}
    \caption{\textit{a)} Experimental setup used for hot-wire anemometry (HWA) measurements over both smooth and rough walls. \textcircled{1} is the upstream pitot tube from which $U_0$ is set, \textcircled{2} shows the 16 pressure taps of the rough wall, \textcircled{3} the NACA 0012 Aerofoil of 1.25 m chord, \textcircled{4} is the location of the drag balance used for skin friction measurements on the rough wall, \textcircled{5} is the traverse to which \textcircled{6} the HWA probe is mounted. \textit{b)} 0.25x0.25 m section of the smooth wall constructed from aluminium sandwich panels. \textit{c)} 0.25x0.25 m section of the rough wall constructed from plywood topped with 3 mm acrylic with 3 mm roughness mounted on top.}
    \label{fig:setup}
\end{figure}

Experiments were carried out in the boundary layer wind tunnel at the University of Southampton, consisting of 5 sections of 2.4 m long, 1 m high and 1.2 m wide. A shallow ramp ($\approx 5^\circ$) is fitted at the start of the test section to remove the step up to the test surface. A 3D turbulator strip of height 0.5mm is placed at the top of the ramp to trip the boundary layer. The tunnel has a turbulence intensity in the central third of the tunnel of 0.6\%. A simplified diagram of the tunnel and the setup from 5 m to 9.8 m is shown in figure \ref{fig:setup}, showing the main elements of the experimental setup. The experiment uses a NACA0012 aerofoil of 1.25 m chord and 1.2 m width mounted on four actuators to adjust the aerofoil's angle of attack. The aerofoil is mounted such leading edge 6.53m from the start of the test section. Both the angle of attack and height of the quarter chord are varied more details of which can be found in section \ref{sec:parameters}. This setup is similar to that of \cite{fritsch_2022} and \cite{vishwanathan_2023_IJHF}, however, we achieve stronger/longer pressure gradient histories due to the length of this wing as well as its location in the freestream. The position of the wing is set by measuring the height of all four corners of the wing above the tunnel floor. This limits the error of setting the wing using only the actuator encoders. 

A pitot tube is mounted one chord upstream of the leading edge of the aerofoil when set to $0^\circ$. The pressure difference is measured using a Furness FCO560 micromanometer. This pitot sets $U_\infty$ throughout this experiment. Temperature and pressure inside the tunnel at the at the exit of the contraction are measured using an RTD TST414 thermometer and Setra 278 barometric pressure transducer, respectively. These are sampled via the tunnel control system after every hot wire point. The tunnel is kept at a constant temperature throughout an experimental run via the tunnel heat exchanger. 

Smooth wall measurements were carried out using aluminium honeycomb sandwich panels, a section of which can be seen in figure \ref{fig:setup}\textit{b}. These are sized for each wind tunnel section to reduce the number of joints and have a thickness of 26.5 mm. For the section where measurements are being taken, the middle half of the tunnel is replaced with 10 mm thick safety glass. Firstly, this reduces the conduction effects of the hot wire close to the tunnel floor. Secondly it allows optical access for wall shear stress measurements. The exception was for the upstream smooth wall, which used the aluminium wall. For rough wall measurements, the floor consists of 15 mm plywood topped with 3 mm PVC onto which an expanded metal mesh is mounted. The roughness runs from the start of the test section for approximately 10.5 m downstream. The metal mesh used has dimensions of 62 mm$\times$30 mm. The longest dimension is in the spanwise direction with a 3 mm height, resulting in an open area of 73\%.  

\subsection{Parameters}\label{sec:parameters}

The main part of this study looks at five different angles of attack: $-8^\circ$, $-4^\circ$, $0^\circ$, $4^\circ$ and $8^\circ$. To examine the effect of PGs measurements are taken at one chord downstream of the trailing edge. For the smooth wall this is $116.1\delta_0$ from the test section start and for the rough wall $55.4\delta_0$. $\delta_0$ is defined as the boundary layer thickness one chord upstream of the aerofoil for a given surface. For the smooth wall, measurements are taken at freestream velocities of 10, 20 and 30 m/s. Rough wall measurements were taken at 10-30 m/s (in steps of 5 m/s). These speeds corresponds to  between \num{6.0e6} $< Re_L <$ \num{19.6e6}. The quarter chord is kept at 0.5 m from the wind tunnel floor for all these cases. Rough wall measurements were taken with the quarter chord at 0.4 m for $-10^\circ$, $-8^\circ$ and $-4^\circ$ at 20, 25 and 30 m/s. Smooth wall data was only taken at this height for the $-10^\circ$ and $-8^\circ$ cases. The reason for this was to provide a greater range of pressure gradient histories and strengths. Measurements were also taken for a height of 0.5 m for $-8^\circ$ and $8^\circ$ one chord upstream of the aerofoil for various speeds mentioned above. The measurment station is $67.9\delta_0$ from the start of the test section for the smooth wall and $32.4\delta_0$ for the rough wall with \num{3.3e6} $< Re_L <$ \num{10.4e6}. ZPG measurements were taken for the rough wall from 15 m/s to 35 m/s (\num{8.9e6} $< Re_L <$ \num{21.0e6}) in steps of 5 m/s. Meanwhile, for the smooth wall, the skin friction and velocity profiles data from \cite{ferreira_2024} is used. This data is taken at the same measurement station as the other data in this work. \cite{ferreira_2024} also has direct skin friction measurements using oil film interferometry (OFI). When plotting different velocities the transparency is reduced when plotting different freestream velocities (i.e increasing Reynolds number). We note that data here captures an extended range of Reynolds numbers that ensures sufficient scale separation. 

\subsection{Pressure Distribution}\label{sec:pressure_method}
Pressure taps are fitted to the wind tunnel floor to measure the wall pressure. For the smooth wall, twenty tubes with an inner diameter of 0.6 mm are fitted to the floor space 0.24 m apart. For rough wall measurements, sixteen pressure taps of 0.5 mm inner diameter were used, spaced approximately 0.265 m apart. Panel method simulations were carried out, suggesting that the upstream and downstream influence of the aerofoil extended one chord. As a result, the taps were placed in this region. This configuration of pressure taps is shown in figure \ref{fig:setup}\textit{a} by the small vertical lines on the underside of the tunnel floor for the rough wall. The mean pressure distribution was recorded using a 64-channel ZOC33/64 Px pressure scanner. The pressure difference was taken referenced to the atmospheric pressure. The pressure data was sampled at multiple points during the hot wire sweep for each pressure gradient case, temperature and pressure data are taken simultaneously.  

\begin{figure}
    \centering
    \includegraphics[width = \textwidth]{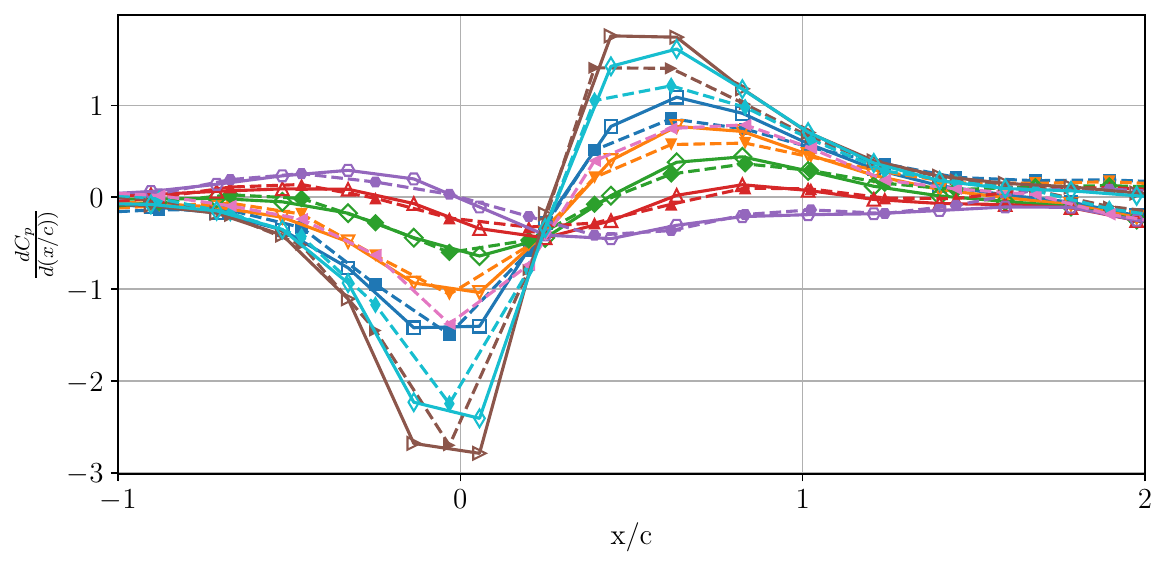}
    \caption{Mean pressure gradient for both smooth and rough walls with respect to $x/c$. For the rough wall cases at $h$ = 0.5m the following symbols are used -  $-8^\circ$: \protect\symbolneight, $-4^\circ$: \protect\symbolnfour  , $0^\circ$: \protect\symbolzero   , $4^\circ$: \protect\symbolfour   and $8^\circ$: \protect\symboleight. For $h$ = 0.4m the rough wall cases are denoted by $-10^\circ$:  \protect\symbolntenlower, $-8^\circ$: \protect\symbolneightlower and  $-4^\circ$: \protect\symbolnfourlower. The ZPG case is given by \protect\symbolzpg. The smooth wall is shown by the same symbols and colours however they are left unfilled. }
        
    \label{fig:dCpdx}
\end{figure}

 Figure \ref{fig:dCpdx} shows the PG history for the different cases tested during this experiment. For the 0.5 m cases, there are two distinct history types. The first are those that have an FPG followed by an APG ($-8^\circ$, $-4^\circ$, $0^\circ$). Secondly, those with an APG followed by an FPG ($4^\circ$ $8^\circ$). All 0.4 m cases fall into the first category. They have greater strength than the 0.5 m cases due to the aerofoil being closer to the wall. The first group of cases will be called APG cases throughout this work. While the second group will be the FPG cases. This is because it is assumed that the pressure gradient type experience second will be more dominant in the resulting boundary layer one chord downstream.
 
 The pressure gradient histories show good agreement between there smooth and rough wall cases. There is some variation in the pressure gradient around the hot wire measurement station. In some cases, the slight FPG comes from the acceleration around the hot wire traverse. In other cases, pressure distribution was taken when the traverse was removed.  The effect of the traverse on the different boundary layers is assumed to be minimal and equal in all cases. Measurements were also taken in a nominal ZPG case, with the wing removed from the wind tunnel. This data was seen to have a slight FPG. The flow accelerates due to the boundary layer growing and the tunnel being a fixed cross-section. 

The cases presented are non-equilibrium pressure distributions since $\beta = (\delta^* / \tau_0) \cdot (dP/dx)$ is not constant. The pressure gradient histories confirm the prediction of the panel method that the influence of the aerofoil extends one chord upstream and downstream of the aerofoil. The boundary layer one chord upstream should be the same since they have had the same pressure gradient history. The hot wire is, therefore, placed one chord downstream of the aerofoil so that the local pressure gradient is the same for all cases. Thus, any difference can be said to be due to the upstream pressure gradient history.  

\subsection{Wall Shear Stress}\label{sec:WSS_method}

When studying turbulent boundary layers, an important quantity is the skin friction; however, the magnitude of these forces is in the order of grams, making it difficult to take direct measurements. OFI is used to measure skin friction directly for smooth wall measurements. Silicon-based oil was used to generate the fringe patterns and imaged using a Lavision ImagerProLX 16MP camera, onto which a Sigma 105mm F2.8 lens is mounted. A Phillips 35W SOX-E bulb provides a single wavelength of light. The wall shear stress is calculated based on the thinning rate of the oil, calculated from the rate of change of the fringe pattern spacing.  

Oil film interferometry is not possible for rough wall cases as it does not provide the true measure of skin friction (in addition to being impossible to implement in the case of current roughness). Therefore, a 0.20$\times$0.20 m drag balance is located in the tunnel floor with the centre of the balance located 8.8m downstream from the start of the test section along the centreline. The balance contains a floating element consisting of a flat plate mounted on a floating element hung from the outer casing with four thin flexures in the corners. The smooth, flat metal plate sits level with the outer casing and wind tunnel floor. Upon this flat plate, the roughness element patch is attached. An electromagnet and distance sensor with a range of $\pm250\mu m$ keeps the displacement at zero when a force is applied. The voltage applied to the electromagnet to maintain zero displacements; for more details, see \cite{ferreira_2024}. Calibration is performed using a known load from 0 to 45g, providing a calibration that can be used to convert the measured voltage during experiments into a drag force.

\subsection{Hot-Wire Anemometry}\label{sec:HWA_method}

Hot-wire anemometry (HWA) is used to acquire velocity profiles from measurements at a single location. A vertical traverse is mounted from the roof as shown in figure \ref{fig:setup}\textit{a} onto which a Dantec 55H21 probe holder is mounted. The traverse is located so that the hot wire sits 9.0m downstream of the start of the test section. The offset between the hot wire and drag balance is 0.1m to prevent any minor interference from the balance.  A single in-house wire probe similar to the Dantec 55P05 probe is used. Consisting of which a $5\mu m$ tungsten wire is soldered; this is then coated with copper, leaving a sensing length of 1mm. This results in a length-to-diameter ratio of 200, meeting the requirement in \citet{Ligrani_1987} that l/d should not be less than 200 to prevent the conduction of the supports affecting the result. The dimensionless wire length $l^+$ given by $(l U_\tau / \nu)$ varies between $21 < l^+ < 74$  for the smooth wall and between $33 < l^+ < 140$ for the rough wall cases. In order to measure the upstream boundary layers, the traverse is moved so that the hot wire probe is mounted at 5.3m from the test section start.  The rest of the setup remains the same as for the downstream cases. 

The overheat ratio was set to 0.8 throughout the experiment.  The sampling time (T) varies from case to case, ensuring at least 19,000 boundary layer turnover times ($TU/\delta$). The signal is read using an NI USB-6212 16-bit DAQ and is also used to read the pitot tube micromanometers. The wire's initial position is calibrated using a microscope camera. The wire is then moved towards the wall to set the initial position. The voltages are adjusted based on deviations from the initial calibration temperature to account for any temperature drift throughout a run. Calibration is carried out with the wing at $0^\circ$ at a height of around 0.55 m from the floor to ensure the wing does not affect the calibration process. The probe is mounted halfway between the boundary layer and the wing. A second pitot is mounted on the traverse at the same height as the probe to calibrate against. Calibration fitting uses a fourth-order polynomial to convert voltages to velocities.  

\section{Results} \label{sec:Results}

This section examines the mean boundary layer velocity profiles from both upstream and downstream of the aerofoil. Direct wall shear stress measurements are also used. The section starts by examining the upstream boundary layer before it experiences the PG history and, subsequently, the downstream location after the flow has experienced different histories of PGs.  

\subsection{Incoming flow}\label{sec:incoming_bl}

The pressure distributions in section \ref{sec:pressure_method} showed a ZPG region approximately one chord upstream of the leading edge. Therefore, it is expected that the incoming boundary layer, one chord upstream of the leading edge, is invariant to the angle of attack. 

\begin{figure}
  \centerline{\includegraphics[width=\textwidth]{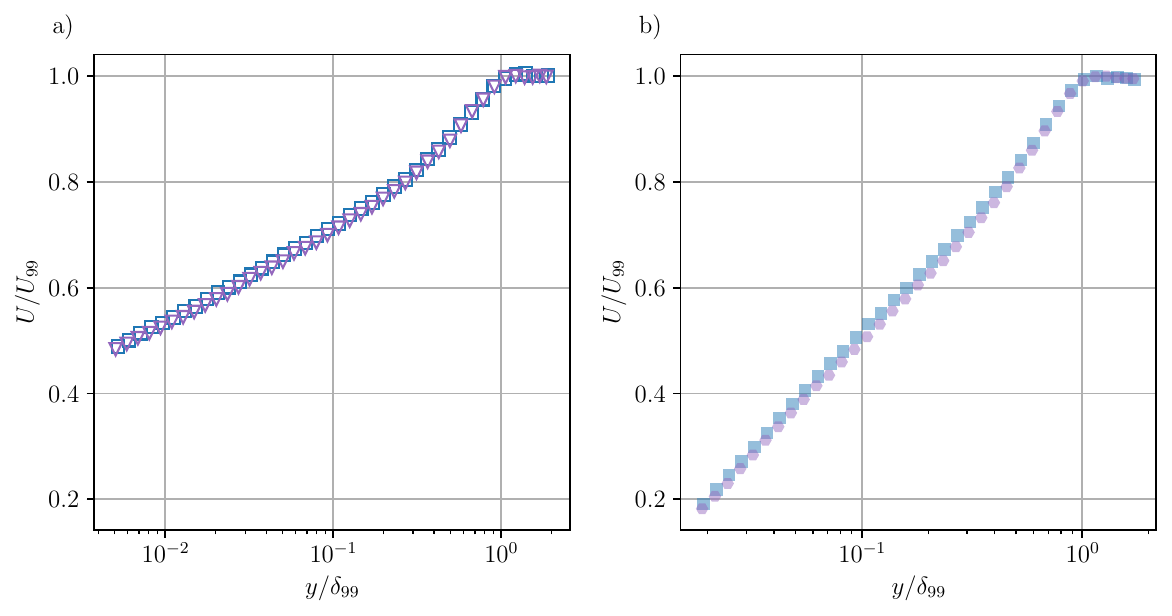}}
  \caption{Mean velocity profiles taken at a hot wire location of 5.3 m from the test section start for $-8^\circ$ and $8^\circ$ with the quarter chord at a height of 0.5 m for (\textit{a}) smooth wall at 30 m/s  (\textit{b}) rough wall at 10 m/s. Symbols and colours are as per figure \ref{fig:dCpdx}.}
\label{fig:upstream_condition}
\end{figure}

\begin{table}
\begin{center}
\def~{\hphantom{0}}
\begin{tabular}{cccccccccc}
Case & Surface & Symbol & $U_\infty$ ($m/s$) & $U_{99}$ ($m/s$) & $\delta$ ($m$) & $\delta^+$ ($m$) & $\theta$ ($m$) & $H$ & $Re_\theta$ \\
\hline
$-8^\circ$ & SW & \symbolneightsw[0.3] & 9.7 & 9.6 & 0.08 &  0.011 & 0.009 & 1.31 & 5612 \\
$-8^\circ$ & SW & \symbolneightsw[0.65] & 19.7 & 18.9 & 0.08 &  0.010 & 0.008 & 1.29 & 10205 \\
$-8^\circ$ & SW & \symbolneightsw[1] & 29.6 & 29.6 & 0.08 &  0.010 & 0.008 & 1.28 & 15864 \\
$-8^\circ$ & RW & \symbolneight[0.3] & 9.9 & 10.0 & 0.16 &  0.033 & 0.021 & 1.58 & 13843 \\
$-8^\circ$ & RW & \symbolneight[0.65] & 19.9 & 20.1 & 0.16 &  0.034 & 0.022 & 1.55 & 29026 \\
$-8^\circ$ & RW & \symbolneight[1] & 29.6 & 29.7 & 0.16 & 0.032 & 0.021 & 1.56 & 40332 \\[.2cm]

$8^\circ$ & SW & \symboleightsw[0.3] & 9.6 & 9.3 & 0.08 &  0.013 & 0.009 & 1.32 & 5902 \\
$8^\circ$ & SW & \symboleightsw[0.65] & 19.4 & 18.7 & 0.08 &  0.011 & 0.009 & 1.30 & 10959 \\
$8^\circ$ & SW & \symboleightsw[1] & 29.0 & 28.7 & 0.08 &  0.011 & 0.009 & 1.29 & 16832 \\
$8^\circ$ & RW & \symboleight[0.3] & 9.9 & 9.6 & 0.16 &  0.036 & 0.022 & 1.60 & 14102 \\
$8^\circ$ & RW & \symboleight[0.65] & 19.9 & 19.9 & 0.17 & 0.036 & 0.023 & 1.57 & 30397 \\
$8^\circ$ & RW & \symboleight[1] & 29.3 & 29.3 & 0.17 &  0.036 & 0.023 & 1.56 & 44975
\end{tabular}  
\caption{Summary of key boundary layer properties for two angles of attack one chord upstream of the aerofoil. Surface given as SW for smooth wall and RW for rough wall.}
  \label{tab:summary_data_upstream}
  \end{center}
\end{table}

The TBL for a smooth wall case, shown in figure \ref{fig:upstream_condition}\textit{a}, exhibits invariance to the angle of attack. Table \ref{tab:summary_data_upstream} shows that the value of the boundary layer thickness ($\delta$) is invariant across all the tested free-stream speeds. The flow over the tested rough wall, as shown in figure \ref{fig:upstream_condition}\textit{b}, exhibits similar results, with the profiles collapsing between two angles of attack (-8$^\circ$ and 8$^\circ$). Table \ref{tab:summary_data_upstream} shows that the boundary layer thickness remains invariant even in the rough wall case. There is a slight variation between different cases; however, $\delta$ is approximately 0.16m in both cases tested. Upstream testing was conducted at two angles of attack: -8$^\circ$ and 8 $^\circ$. Since no dependence was observed at these extreme angles, it can be reasonably assumed that intermediate angles will show no dependence. This result validates the assumption made during the planning stage that the relevant pressure gradient history extends from one chord length upstream to one chord length downstream.

\subsection{Mean Flow - After experiencing pressure gradient history}\label{sec:mean_flow}

\begin{figure}
    \centering
    \includegraphics[width=1\linewidth]{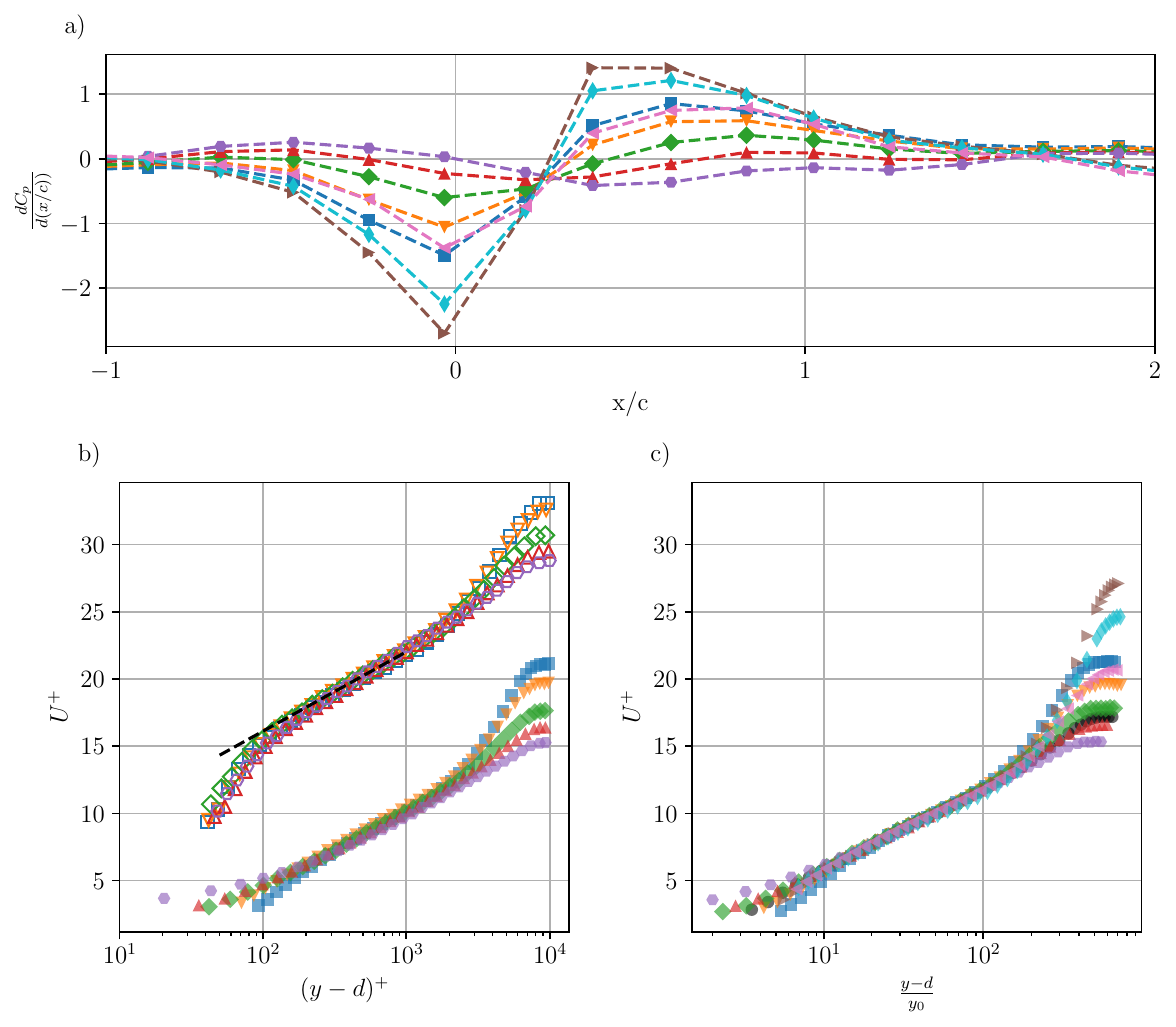}
    \caption{(\textit{a}) Cut down version of figure \ref{fig:dCpdx} repeated to aid interpretation showing $dCp/d(x/c)$ variation for rough wall cases (\textit{b}) Inner scaled velocity profiles at $Re_\tau \approx 6800-8300$ for both smooth and rough wall cases at 0.5 m. The dashed black line shows the log region from \ref{eq:complete_velocity_profile_SW}. (\textit{c}) Rough wall velocity profiles for 20 m/s for the 0.4 m, 0.5 m and ZPG cases. In both plots, $d$ is the zero plane displacement, which for a smooth wall is zero. The x axis is scaled using $y_0$, this results in the collapse of the log region of the profiles. Symbols and colours are as per figure \ref{fig:dCpdx}.}
    \label{fig:velocity_profiles}
\end{figure}

\begin{table}
\begin{center}
\begin{tabular}{ccccccccccccc}
Case & Surface & Symbol& \begin{tabular}[c]{@{}c@{}}$U_\infty$\\  ($m/s$)\end{tabular} & \begin{tabular}[c]{@{}c@{}}$U_{99}$ \\ ($m/s$)\end{tabular} & \begin{tabular}[c]{@{}c@{}}$U_\tau$ \\ ($m/s$)\end{tabular} & \begin{tabular}[c]{@{}c@{}}$\delta$\\  ($m$)\end{tabular} & \begin{tabular}[c]{@{}c@{}}$\delta^*$ \\ ($m$)\end{tabular} & \begin{tabular}[c]{@{}c@{}}$\theta$ \\ ($m$)\end{tabular} & $H$ & $Re_\tau$ & $Re_\theta$ & $\Pi$ \\

\hline
ZPG & SW & \symbolzpgsw[0.3] & 9.8 & 10.0 & 0.35 & 0.12 &  0.017 & 0.013 & 1.30 & 2758 & 8874 & 0.52 \\
ZPG & SW &\symbolzpgsw[0.65]  & 19.9 & 19.9 & 0.68 & 0.12 &  0.016 & 0.013 & 1.28 & 5520 & 17164 & 0.52 \\
ZPG & SW & \symbolzpgsw[1] & 29.5 & 30.0 & 0.98 & 0.12 &  0.015 & 0.012 & 1.26 & 7636 & 24281 & 0.52 \\
ZPG & RW & \symbolzpg[0.65] & 19.9 & 19.8 & 1.15 & 0.22 &  0.044 & 0.029 & 1.51 & 17001 & 38146 & 0.29 \\
ZPG & RW & \symbolzpg[1]  &29.9 & 29.7 & 1.72 & 0.23 &  0.044 & 0.029 & 1.50 & 26074 & 57530 & 0.29 \\[.2cm]
$-8^\circ$ (0.5m) & SW & \symbolneightsw[0.3]  &9.5 & 10.0 & 0.31 & 0.14 &  0.026 & 0.018 & 1.43 & 2907 & 12094 & 1.26 \\
$-8^\circ$ (0.5m) & SW & \symbolneightsw[0.65]  &19.3 & 20.3 & 0.62 & 0.14 &  0.024 & 0.017 & 1.39 & 5837 & 23783 & 1.26 \\
$-8^\circ$ (0.5m) & SW & \symbolneightsw[1] & 28.7 & 30.2 & 0.91 & 0.14 &  0.023 & 0.017 & 1.38 & 8368 & 34796 & 1.26 \\
$-8^\circ$ (0.5m) & RW & \symbolneight[0.3]  &9.9 & 10.4 & 0.50 & 0.23 & 0.063 & 0.037 & 1.72 & 7830 & 25177 & 1.18 \\
$-8^\circ$ (0.5m) & RW &  \symbolneight[0.65] & 20.1 & 21.7 & 1.02 & 0.25 &  0.065 & 0.038 & 1.71 & 16592 & 53984 & 1.18 \\
$-8^\circ$ (0.5m) & RW &  \symbolneight[1] & 30.1 & 32.5 & 1.51 & 0.26 &  0.065 & 0.038 & 1.71 & 25452 & 81527 & 1.18 \\[.2cm]
$-4^\circ$ (0.5m) & SW & \symbolnfoursw[0.3] & 9.6 & 10.0 & 0.33 & 0.13 &  0.022 & 0.016 & 1.38 & 2811 & 10803 & 1.03 \\
$-4^\circ$ (0.5m) & SW & \symbolnfoursw[0.65] & 19.5 & 20.3 & 0.65 & 0.13 &  0.022 & 0.016 & 1.36 & 5673 & 21664 & 1.03 \\
$-4^\circ$ (0.5m) & SW & \symbolnfoursw[1]  &28.9 & 30.5 & 0.94 & 0.13 &  0.020 & 0.015 & 1.36 & 7942 & 30326 & 1.03 \\
$-4^\circ$ (0.5m) & RW & \symbolnfour[0.3] & 9.9 & 10.4 & 0.53 & 0.22 &  0.053 & 0.032 & 1.63 & 7787 & 21991 & 0.81 \\
$-4^\circ$ (0.5m) & RW & \symbolnfour[0.65] & 20.2 & 21.6 & 1.10 & 0.25 &  0.055 & 0.034 & 1.63 & 17446 & 47442 & 0.81 \\
$-4^\circ$ (0.5m) & RW & \symbolnfour[1] & 29.9 & 31.8 & 1.61 & 0.24 &  0.056 & 0.034 & 1.64 & 25484 & 71023 & 0.81 \\[.2cm]
$0^\circ$ (0.5m) & SW & \symbolzerosw[0.3] & 9.6 & 10.1 & 0.35 & 0.12 &  0.019 & 0.014 & 1.35 & 2780 & 9495 & 0.73 \\
$0^\circ$ (0.5m) & SW & \symbolzerosw[0.65] & 19.5 & 20.1 & 0.67 & 0.12 & 0.018 & 0.013 & 1.33 & 5325 & 18122 & 0.73 \\
$0^\circ$ (0.5m) & SW & \symbolzerosw[1] & 28.9 & 30.1 & 0.98 & 0.11 &  0.017 & 0.013 & 1.31 & 7558 & 25698 & 0.73 \\
$0^\circ$ (0.5m) & RW & \symbolzero[0.3] & 10.0 & 10.3 & 0.59 & 0.21 &  0.045 & 0.029 & 1.57 & 8049 & 19063 & 0.48 \\
$0^\circ$ (0.5m) & RW & \symbolzero[0.65] & 20.0 & 21.1 & 1.19 & 0.21 &  0.045 & 0.029 & 1.57 & 16974 & 40024 & 0.48 \\
$0^\circ$ (0.5m) & RW & \symbolzero[1] & 29.9 & 31.7 & 1.77 & 0.22 &  0.046 & 0.029 & 1.58 & 25672 & 61189 & 0.48 \\[.2cm]
$4^\circ$ (0.5m) & SW & \symbolfoursw[0.3] & 9.5 & 10.1 & 0.37 & 0.11 &  0.015 & 0.012 & 1.32 & 2741 & 8015 & 0.48 \\
$4^\circ$ (0.5m) & SW & \symbolfoursw[0.6] & 19.3 & 20.5 & 0.71 & 0.11 &  0.014 & 0.011 & 1.30 & 5110 & 15553 & 0.48 \\
$4^\circ$ (0.5m) & SW & \symbolfoursw[1] & 28.8 & 30.4 & 1.03 & 0.10 &  0.014 & 0.011 & 1.29 & 7347 & 22243 & 0.48 \\
$4^\circ$ (0.5m) & RW & \symbolfour[0.3] & 9.9 & 10.3 & 0.63 & 0.20 &  0.038 & 0.025 & 1.52 & 8335 & 16839 & 0.23 \\
$4^\circ$ (0.5m) & RW & \symbolfour[0.65] & 20.2 & 21.6 & 1.30 & 0.21 &  0.039 & 0.026 & 1.52 & 17447 & 36034 & 0.23 \\
$4^\circ$ (0.5m) & RW & \symbolfour[1] & 29.9 & 31.8 & 1.91 & 0.20 &  0.039 & 0.025 & 1.54 & 26131 & 53378 & 0.23 \\[.2cm]
$8^\circ$ (0.5m) & SW & \symboleightsw[0.3] & 9.5 & 10.5 & 0.40 & 0.10 &  0.013 & 0.010 & 1.29 & 2679 & 7053 & 0.28 \\
$8^\circ$ (0.5m) & SW & \symboleightsw[0.65] & 19.2 & 20.9 & 0.76 & 0.10 & 0.012 & 0.010 & 1.27 & 5087 & 13566 & 0.28 \\
$8^\circ$ (0.5m) & SW & \symboleightsw[1] & 28.7 & 30.9 & 1.08 & 0.09 &  0.011 & 0.009 & 1.28 & 6893 & 18510 & 0.28 \\
$8^\circ$ (0.5m) & RW & \symboleight[0.3] & 10.1 & 10.7 & 0.70 & 0.18 &  0.032 & 0.021 & 1.50 & 8090 & 14607 & 0.04 \\
$8^\circ$ (0.5m) & RW & \symboleight[0.65] & 20.1 & 22.2 & 1.44 & 0.19 & 0.034 & 0.023 & 1.50 & 17803 & 32851 & 0.04 \\
$8^\circ$ (0.5m) & RW & \symboleight[1] & 30.0 & 32.9 & 2.12 & 0.20 &  0.035 & 0.023 & 1.50 & 27423 & 50254 & 0.04 \\[.2cm]
$-10^\circ$ (0.4m) & SW & \symbolntenlowersw[0.3] & 9.5 & 10.3 & 0.30 & 0.16 &  0.034 & 0.022 & 1.54 & 3147 & 15445 & 1.84 \\
$-10^\circ$ (0.4m) & SW & \symbolntenlowersw[0.65] & 19.6 & 21.4 & 0.61 & 0.16 &  0.032 & 0.022 & 1.48 & 6418 & 31050 & 1.84 \\
$-10^\circ$ (0.4m) & SW & \symbolntenlowersw[1] & 29.5 & 32.4 & 0.89 & 0.16 &  0.030 & 0.021 & 1.45 & 9466 & 44663 & 1.84 \\
$-10^\circ$ (0.4m) & RW & \symbolntenlower[0.65] & 20.0 & 22.2 & 0.82 & 0.29 &  0.092 & 0.049 & 1.89 & 15761 & 71538 & 2.32 \\
$-10^\circ$ (0.4m) & RW & \symbolntenlower[1] & 30.2 & 33.6 & 1.22 & 0.30 &  0.093 & 0.049 & 1.90 & 23646 & 107286 & 2.32 \\[.2cm]
$-8^\circ$ (0.4m) & SW & \symbolneightlowersw[0.3] & 9.5 & 10.1 & 0.30 & 0.14 &  0.029 & 0.019 & 1.48 & 2934 & 13285 & 1.55 \\
$-8^\circ$ (0.4m) & SW & \symbolneightlowersw[0.65] & 19.3 & 20.8 & 0.61 & 0.15 &  0.028 & 0.019 & 1.44 & 6133 & 27425 & 1.55 \\
$-8^\circ$ (0.4m) & SW & \symbolneightlowersw[1] & 28.7 & 30.8 & 0.89 & 0.15 & 0.026 & 0.018 & 1.41 & 8864 & 38399 & 1.55 \\
$-8^\circ$ (0.4m) & RW & \symbolneightlower[0.65] & 19.9 & 21.4 & 0.87 & 0.28 &  0.081 & 0.045 & 1.78 & 16068 & 64203 & 1.78 \\
$-8^\circ$ (0.4m) & RW & \symbolneightlower[1] & 30.3 & 32.9 & 1.33 & 0.28 &  0.080 & 0.045 & 1.76 & 24052 & 95654 & 1.78 \\[.2cm]
$-4^\circ$ (0.4m) & RW & \symbolnfourlower[0.65] & 19.9 & 21.1 & 1.02 & 0.25 &  0.058 & 0.036 & 1.61 & 16494 & 50287 & 0.99 \\
$-4^\circ$ (0.4m) & RW & \symbolnfourlower[1] & 30.1 & 31.9 & 1.54 & 0.25 &  0.061 & 0.037 & 1.65 & 24989 & 76944 & 0.99

\end{tabular}

  \caption{Summary of hot wire data taken 9.03m from the inlet of the wind tunnel for different pressure gradient histories.}
  \label{tab:summary_data}
\end{center}

\end{table}

The velocity profiles of the rough and smooth wall TBLs are inner scaled using the directly measured friction velocity. Figure \ref{fig:velocity_profiles}\textit{b} shows the inner scaled profiles at $Re_\tau \approx 8000$, i.e. freestream speeds of $30$ m/s for the smooth wall and $10$ m/s for the rough wall cases. Table \ref{tab:summary_data} gives the variation of the main boundary layer parameters used throughout the investigation. For the smooth wall case, $Re_\tau$ ranges from 8310 for the $-8^\circ$ to 6900 for the $8^\circ$. The variation in $Re_\tau$ is much lower for the rough wall, with values varying from 7830 to 8330 for the $-8^\circ$ and $4^\circ$, respectively. 

The boundary layer profiles of the flow over the smooth wall collapse into the log region. It is seen that any case with APG just upstream of the measurement location results in a larger wake region (i.e. larger $\Pi$) and earlier deviation from the log law region. As the angle of attack increases, the resulting pressure gradient just upstream of the measurement location becomes more favourable, the wake becomes smaller (i.e. smaller $\Pi$) and the log region extends further away from the wall. The rough wall cases show the same wake and log region trends. In figure \ref{fig:velocity_profiles}\textit{b}, there is a clear downward shift of the profiles from the smooth wall cases due to the roughness effects. As explained in section \ref{sec:introduction}, this is because of the extra momentum loss and increased drag which depends on the type of roughened surface. 

\begin{table}
\begin{center}
\def~{\hphantom{0}}
\begin{tabular}{cccccccccc}
 & -10$^\circ$ & -8$^\circ$ & -4$^\circ$ & 0$^\circ$ & 4$^\circ$ & 8$^\circ$ \\ 
 $d / d_{ZPG} $ (0.5 m cases) & & 0.09 & 0.69 & 1.37 & 1.56 & 2.19 \\
$d / d_{ZPG} $ (0.4 m cases) & 0.35 & 0.00 & 0.00 &  & &  \\

$y_0 / y_{0_{ZPG}} $ (0.5 m cases) & & 1.17 & 1.07 & 1.05 & 1.03 & 1.05 \\
$y_0 / y_{0_{ZPG}} $ (0.4 m cases) & 0.98 & 0.94 & 0.97 &  & &  \\

\end{tabular}
\caption{Values of $d / d_{ZPG}$ and $y_0 / y_{0_{ZPG}}$ for different pressure gradient histories with values of $d_{ZPG} = 0.00137m$ and $y_{0_{ZPG}} = 0.000457m$.}
\label{tab:log_law_parameters}
\end{center}
\end{table}

The roughness length scale chosen throughout this work is $y_0$. It was chosen since all the flow measurements were taken within the fully-rough regime, as was shown by the skin friction measurements. In equation \ref{eq:complete_velocity_profile_RW}, the two unknowns in the log region are $d$ (the zero plane displacement) and $y_0$ (the roughness length scale). Using the measurement value of skin friction, we first fit the zero plane displacement while ensuring $d$ should be less than the roughness height, $k$ (\cite{castro_vanderwel}). The value of $d$ is fitted using the diagnostic function, $\Xi = \frac{1}{U_\tau} \frac{dU}{dy} (y - d)$, which is equal to $1/\kappa$ in the log region. The value of d is chosen to give the longest log region possible within the acceptable error range. The acceptable error range is defined such that the average deviation of the points chosen to be in the log region is less than $\pm5\%$ from $1/\kappa$. The resultant values of zero plane displacement are shown in table \ref{tab:log_law_parameters}. The results show APG just upstream of the measurement location reduces the value of the zero plane displacement, and FPG increases it. For the strong APG cases, the value of zero plane displacement is approximately zero. This suggests that the range of $d$ values chosen as bounds may be limiting and that $d$ could well be negative; however, this would be inconsistent with the recommendations from previous work. Further work may be necessary to examine the influence of pressure gradients on this quantity. Since most of our work involves high Reynolds numbers, the exact choice of $d$ has minimal impact on the results in the following sections, so we opt to leave it unchanged. The value of $d_{ZPG}$ was found to be 0.46$k$ and this is consistent with the work of \cite{squire_2016}, who, for ZPG flows, suggested choosing $d$ as $k/2$.

The method for finding $d$ defines the bounds of the log region where the error remains within an acceptable range. Within this region, we obtain $y_0$ as an offset from the smooth wall. The results are shown in table \ref{tab:log_law_parameters}. While for the zero plane displacement, there is a trend shown with pressure gradient history, for the roughness length scale, there is no clear trend. The $-8^\circ$ case shows a small increase in the roughness length scale compared to the ZPG value. However, the cases at 0.4 m have roughness length scale values lower than the ZPG case. The absence of a clear trend with the pressure gradient history, along with the minimal variation in values, suggests that $y_0$ is unaffected by the flow history. Any differences are attributed to the fitting process and the selected boundary layer region. Especially for high APG cases just upstream of the measurement location where the log region is small. This means that some of the wake region will likely be fitted to the log region, thus affecting the value of $y_0$. Regardless, the maximum deviation in $y_0$ across the different cases is less than 20\% and, in fact, less than 10\% for the majority of cases examined here. This is much smaller than the deviations reported in \cite{vishwanathan_2023_IJHF} and is presumably because of the scale separation that was achieved in this study where a considerable log region can be identified across all profiles. Moreover, an independent measure of $U_\tau$ limits the uncertainty in fitting leading to better estimates of $y_0$.

The inner scale velocity profiles as a function $(y-d)/y_0$ is shown in figure \ref{fig:velocity_profiles}\textit{c}. This scaling provides a good collapse for all velocity profiles in the log region. The scatter between the different cases in the log region is reduced compared to the rough wall cases in figure \ref{fig:velocity_profiles}\textit{b}. All profiles also exhibit a clear wake region that changes with the nature of the pressure gradient just upstream of the measurement location. However, it should be noted that this local wake is an integral effect of the entire pressure gradient history experienced by the flow. It is shown that the profile of the ZPG case has a wake between those at $0^\circ$ and $4^\circ$ at 0.5 m. This is expected because the type of pressure gradients reverses order between these two cases. It can be seen that the $-4^\circ$ at 0.4 m causes a larger wake than at 0.5 m but it is smaller than that is seen at $-8^\circ$ case at 0.5 m. This pattern in the velocity profiles follows the same order seen in figure \ref{fig:dCpdx} between different cases. There is clearly a complex relationship between the wake profile and the imposed pressure gradient history. Be that as it may, the correlation between $C_f$ and wake is consistent regardless of the history for both smooth and rough walls. 

\begin{figure}
    \centering
    \includegraphics[width=1\linewidth]{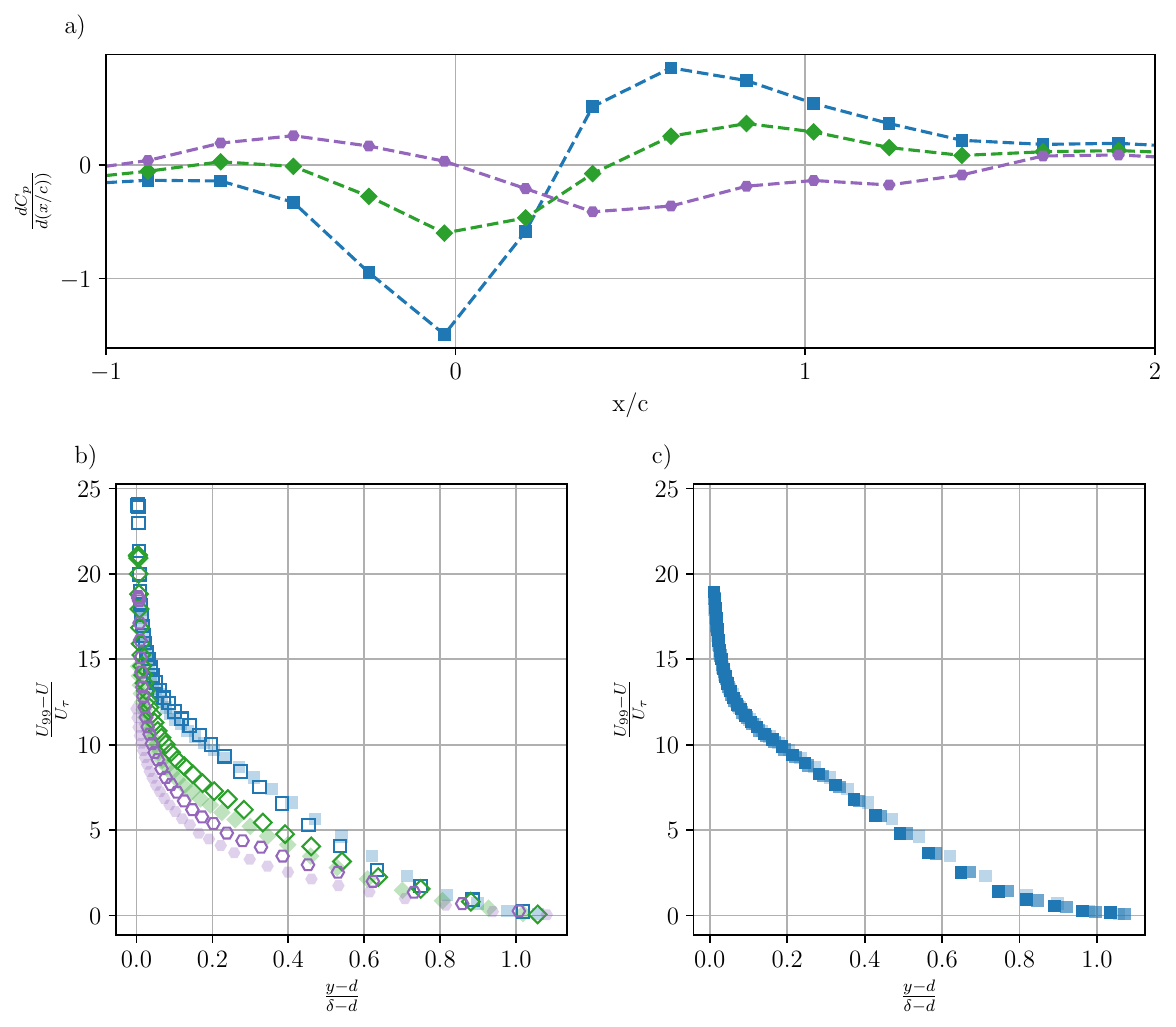}
    \caption{(\textit{a}) Cut down version of figure \ref{fig:dCpdx} repeated to aid interpretation showing $dCp/d(x/c)$ for the $-8^\circ$, $0^\circ$ and $8^\circ$ at 0.5 m for the rough wall (\textit{b}) Comparison of the velocity deficit profiles for $-8^\circ$, $0^\circ$ and $8^\circ$ at 0.5 m for both smooth and rough walls. (\textit{c}) shows the variation in velocity deficit profile for rough wall with Reynolds number for $-8^\circ$ at a height of 0.5 m for 10, 20 and 30 m/s. In both plots, $d$ is the zero plane displacement, which is zero for a smooth wall. Symbols and colours are as per figure \ref{fig:dCpdx}.}
    \label{fig:deficit_profiles}
\end{figure}

The velocity deficit profiles enable the examination of the outer wake in more detail. The results for three angles of attack are shown in figure \ref{fig:deficit_profiles}\textit{b}. For the $-8^\circ$ case, there is a good collapse of the profiles between the smooth and rough wall cases. This would suggest that the integral effect of the pressure gradient and roughness on the outer region is similar to that of the smooth wall for this combination of pressure gradient history. However, no collapse occurs as the pressure gradient becomes more favourable immediately upstream of the measurement location. There is no outer layer similarity because the boundary layer growth of the rough wall is larger than that of the smooth wall. Therefore, the integral effect of roughness and pressure gradient between the smooth and rough walls are not consistent. As a result, the rough wall in the presence of an FPG just upstream of the measurement location has smaller wake strengths (lower $\Pi$) than that of a smooth wall. Therefore, it may not be trivial to have information for a smooth wall with a given pressure gradient history (even at similar $Re_\tau$ and identical local $\beta$) and use that to infer properties of a rough wall. As suggested by \cite{volino_2023} and \cite{vishwanathan_2023_IJHF} it appears important to match the $\beta$ history to obtain complete similarity but that is almost impossible to devise in experiments (since $\beta$ is an output while $dP/dx$ is the only input). Therefore, we need new relationships that will allow us to infer information about these flows based on local measurements. 

Figure \ref{fig:deficit_profiles}\textit{c} shows that deficit profiles (for $-8^\circ$) collapse across different Reynolds numbers, and similar trends are observed for the other angles of attacks across smooth and rough wall cases. Based on this observation, the wake parameter, $\Pi$ for each case is calculated using all available velocities for a given angle of attack. The fitting is carried out using equation \ref{eq:deficit_profile}, which only depends on directly measured values. The results of this fit are seen in table \ref{tab:summary_data} in the column labelled $\Pi$. The values of $\Pi$ obtained with the fitting process confirm that TBLs under APG just upstream of the measurement location have larger wake strengths compared to ZPG flows. In contrast, FPGs reduce the wake strength. As shown in the deficit profiles, the wake values of the APG cases are similar. The variation in $\Pi$ shows some interesting trends. Firstly, for 0.5 m cases, the smooth wall wake strength is always greater than the rough wall. This is despite the similar pressure gradient histories shown in figure \ref{fig:dCpdx}. As explained previously, this is due to the difference in boundary layer thicknesses and the resulting acceleration of the flow. Further evidence of this is that for the ZPG TBLs, the wake strength is much higher for the smooth wall, suggesting an FPG effect.  

The trends observed for the case where the wing is mounted at a distance of 0.5 m far from the wall do not occur at 0.4 m. With the aerofoil mounted closer to the wall, and so with stronger APG conditions, the smooth wall TBL exhibits a lower wake strength if compared to the rough wall velocity profiles. As shown in figure \ref{fig:dCpdx}, the match between smooth and rough wall cases worsens as the pressure gradient strength increases. However, the difference suggests that the smooth wall has stronger peak pressure gradients. Therefore, one might expect the smooth wall to have a larger wake due to the APG; however, this is not what is observed. One possible explanation is that a thicker boundary layer is more susceptible to APG rather than FPG and thus results in stronger wake strength. 

\subsection{Skin Friction - After experiencing pressure gradient history}\label{sec:skin_friction}

\begin{figure}
  \centerline{\includegraphics[width=\textwidth]{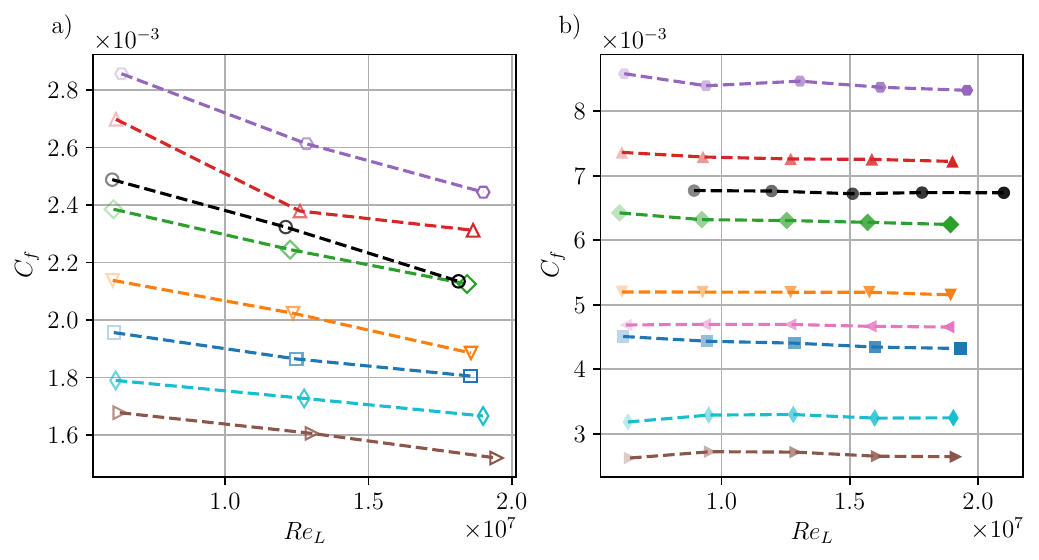}}
  \caption{Skin friction coefficient one chord downstream of the trailing edge of the aerofoil for both 0.4 m and 0.5 m cases. (\textit{a}) Skin friction coefficient for a smooth wall and (\textit{b}) skin friction coefficient for a rough wall. Symbols and colours are as per figure \ref{fig:dCpdx}.}
\label{fig:skin_friction_coefficent}
\end{figure}

Most of the previous experimental studies on pressure gradient effects on TBLs over both smooth and rough walls had inferred skin friction from the velocity profiles. These methods introduce uncertainty into the measurements. Therefore, in this presented work, we aim to improve the experimental investigation by directly measuring the wall shear stress, as outlined in section \ref{sec:WSS_method}. The skin friction coefficient for the smooth wall is shown in figure \ref{fig:skin_friction_coefficent}\textit{a}. As expected for the ZPG smooth wall, the skin friction coefficient reduces as the Reynolds number increases (\cite{schultz_2013}). The $8^\circ$ angle exhibits the strongest favourable pressure gradient just upstream of the measurement location and therefore exhibits the highest skin friction. Conversely, the $-8^\circ$ angle is characterised by the strongest APG just upstream of the measurement location and hence the lowest skin friction. The other angles are arranged in order of increasing angle of attack between these two cases. The ZPG case lies between the $0^\circ$ and $4^\circ$ cases. From figure \ref{fig:dCpdx} this might be expected since these are the mildest two cases. $0^\circ$ case experiences a mild APG upstream of the hot wire while the $4^\circ$ experiences an FPG region. Therefore, it makes sense that the ZPG cases fit between these two cases. These results indicate that the immediate upstream pressure gradient is more critical than those further upstream. This would indicate that any model that includes history effects should account for this variation in importance. 

The rough wall skin friction coefficients, in figure \ref{fig:skin_friction_coefficent}\textit{b}, do not depend on $Re_x$, meaning that the flow is in the fully-rough regime. The order of cases observed with the smooth wall is replicated with the rough wall at the 0.5 m height for the angles tested. The ZPG case follows this trend, positioning between the $0^\circ$ and $4^\circ$ cases. The rough wall cases at 0.4 m show similar trends with decreasing skin friction as the angle of attack becomes increasingly negative. As expected, the $-4^\circ$ and $-8^\circ$ cases at 0.4 m have lower skin friction than the equivalents at 0.5 m since the aerofoil is closer to the surface and thus has a stronger pressure gradient history. The local skin friction measurement clearly retains the history of the PG type and strength. Furthermore the PG type in second half of the domain is more dominant in overall local skin friction than the PG type in the first half of the domain.

\subsection{Skin friction estimation from mean flow}

\begin{figure}
    \centering
    \includegraphics[width=\textwidth]{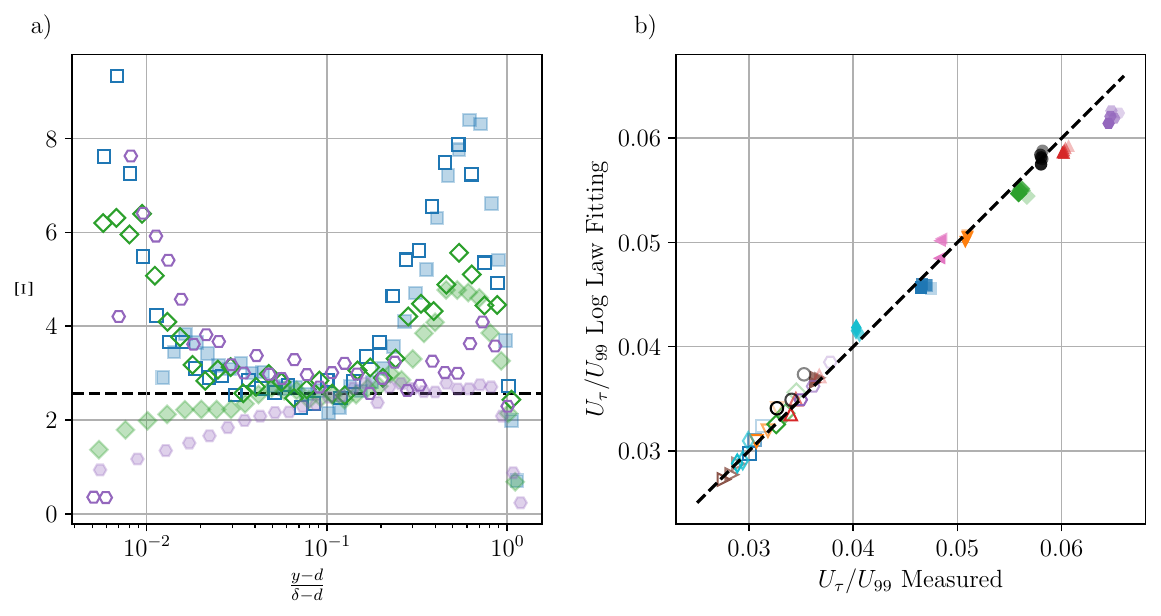}
    \caption{(\textit{a}) Diagnostic plot showing $\Xi = (y-d) \cdot (dU^+/dy)$ for $-8^\circ$, $0^\circ$ and $8^\circ$, both smooth and rough wall are shown. The black dashed line shows $1/\kappa$. (\textit{b}) Comparison of $U_\tau/U_{99}$ from log law fitting vs $U_\tau/U_{99}$ from direct measurement techniques for both smooth and rough walls. The black dashed line is that of $y = x$, which would be true for a perfect prediction from in-direct methods. Symbols and colours are as per figure \ref{fig:dCpdx}.}
    \label{fig:log_law_vs_direct}
\end{figure}

Direct skin friction measurements are rare, with a majority of previous studies relying on the mean velocity profile to estimate the friction velocity and, therefore, the skin friction coefficient. With the data presented in the previous sections, it is possible to discern the difference between skin friction estimation using mean flow and direct skin friction measurements. The log law fitting method for the smooth wall is a simple problem using log region in equation \ref{eq:complete_velocity_profile_SW}, where the only unknown from the raw data is $U_\tau$. For the rough wall cases, we will use the log region in equation \ref{eq:complete_velocity_profile_RW} assuming $y_{0_{ZPG}}$ to be the $y_0$ for all cases due to the very small variation. Therefore, when fitting the log law to this section, the zero-plane displacement was fixed at half the roughness height (0.5$k$ = 1.5 mm), following the method used by \cite{squire_2016}. This approach leaves only one unknown parameter, $U_\tau$, which can be determined through curve fitting. Figure \ref{fig:log_law_vs_direct}\textit{a} shows the diagnostic function, defined as $\Xi = (y-d) \cdot (dU^+/dy)$. The black dashed line represents $1/\kappa$, as $\Xi$ is equal to this value in the log region, as indicated by equations \ref{eq:complete_velocity_profile_SW} and \ref{eq:complete_velocity_profile_RW}. This allows the extent of the log law to be assessed by examining deviations from $1/\kappa$. It can be seen that the smooth wall has an earlier departure from the log than the rough wall cases. Therefore, for the smooth wall cases, the log law is fitted with a minimum of seven points up to $0.15\delta$, while for the rough wall cases, this is extended up to $0.2\delta$.

Overall, figure \ref{fig:log_law_vs_direct}\textit{b} shows good agreement between the direct measurement techniques and the predicted value from the mean flow profiles. For the smooth wall cases, the error varies between 0.1\% and 4.2\%, while for rough wall cases, the error is between 0.6\% and 4.8\%. The largest percentage error for the smooth wall cases is seen for the 10 m/s cases. This was expected since the values of $U_\tau$ are the smallest. The largest error for the rough wall cases is seen in the FPG cases where the log law fitting method underpredicts the skin friction. These boundary layers have the longest log region and highest skin friction. Despite having a longer log region, the error can be attributed to the fitting process. For consistency, the log region is assumed to end at $0.2\delta$; however, as seen in figure \ref{fig:log_law_vs_direct}\textit{a}, the log region can be said to extend well into the outer region of the flow. Smooth wall data is expected to show good agreement since there is only one unknown in the fitting problem. For rough walls, the scatter is minimal since the $y_0$ is assumed a priori for a given surface. Otherwise, the scatter would be significantly greater as both $y_0$ and $U_\tau$ would need to be fitted simultaneously and are interdependent. We note that the work here shows that fitting methods do indeed work where the flow has locally reached a zero pressure gradient (i.e. relaxing flows). However, it is unclear if this is still the case when the flow is locally subjected to a pressure gradient and that requires further work. 

\section{Development of a correlation model for skin friction} \label{sec:Models}

Examination of mean flow characteristics indicated that the pressure gradient just upstream of the measurement location was important. The data showed the wake strength of smooth and rough walls with the same input pressure gradients are not similar. However, the roughness length in the measurement location did not exhibit any history effects and that suggests that any changes previously reported would have been due to a lack of scale separation or direct skin friction measurements. Finally, the data also showed that there is a clear correlation between local skin friction and local wake strength regardless of pressure gradient history. Therefore, it may be possible to develop a correlation model for skin friction and wake strength with some further modelling assumptions. This is explored further in this section. 

We take inspiration from work of \citet{vinuesa_2017} who showed that for APG flows, the local skin friction can be predicted based on the skin friction ($C_f$) and shape factor ($H = \delta^*/\theta$) of ZPG flows,  and the streamwise-averaged pressure gradient parameter ($\overline{\beta}$). They showed that for weak distribution of $\beta$ (i.e. ranging from $0\sim2$), 

\begin{equation}
C_f^{APG} = \frac{C_f^{ZPG}}{(H^{ZPG})^{\overline{\beta}/2}}~~{\rm where}~~\overline{\beta} = \frac{1}{x_{DS} - x_{US}}\int_{x_{US}}^{x_{DS}} \beta(x) dx \label{betabar}
\end{equation}

Here, the $\overline{\beta}$ is the streamwise-averaged pressure gradient parameter, averaged from an upstream streamwise location, $x_{US}$, up to a downstream streamwise location, $x_{DS}$ (from ${Re_\theta}_{US}$ to ${Re_\theta}_{DS}$). It is unclear if this $\overline{\beta}$ will be feasible for flows that experience both APG and FPG (or vice versa) in succession as well as for flows with surface roughness. For experiments it is often impracticable to obtain complete $\beta$ history due to the due to the need for many streamwise measurement stations (either for smooth or rough walls). Moreover, it is unclear if the streamwise average as proposed in equation \ref{betabar} is sufficient to capture history effects. The previous section showed that the locations just upstream of a given point are more important than locations further upstream. Therefore, we need to revise the approach to get better skin friction models for arbitrary pressure gradient histories. To tackle this challenge, we evaluate the mean velocity profile relationships in \ref{eq:complete_velocity_profile_SW} and \ref{eq:complete_velocity_profile_RW} at $y = \delta$, to get $U^+_{99}$ and this is directly related to $C_f$ (= $2/{U^+_{99}}^2$). 

\begin{equation}
    \sqrt{\frac{2}{C_f^{PG}}} - \sqrt{\frac{2}{C_f^{ZPG}}} = \frac{1}{\kappa} ln \left(\frac{y_0^{ZPG}}{y_0^{PG}}\right) + \frac{2}{\kappa} (\Pi^{PG} - \Pi^{ZPG}) + \frac{1}{\kappa} ln\left(\frac{\delta_{PG}^+}{\delta_{ZPG}^+}\right)
    \label{eq:predict_cf_diff}
\end{equation}

Here, the superscript $PG$ refers to an arbitrary pressure gradient case and $ZPG$ is the zero-pressure-gradient case. Here, $\delta^+ = Re_\tau = U_\tau \delta/\nu$ is the friction-velocity-based Reynolds number, and the final term will be zero if we match the Reynolds numbers between PG and ZPG cases. This equation only depends on $y_0$ and $\Pi$ for matched $Re_\tau$ cases. If $y_0$ does not change with pressure gradient as established in section \ref{sec:Results}, then, the change in $C_f$ is entirely due to changes in $\Pi$. It should be noted that the work of \cite{castro_2007} developed similar correlations to obtain the skin friction of rough walls flows at ZPG conditions where he showed that $C_f = f(\theta/y_0, H, \Pi)$. In fact, the variation in local $\Pi$ accounts for flows that do not satisfy outer-layer similarity. However, it is possible to interpret the relationship to be an effect of external pressure gradient history. In that case, the correlations in \cite{castro_2007} are analogous to the relationship in \ref{eq:predict_cf_diff}. Both require knowledge of the local value of $\Pi$ (in addition to the value of $H$ and $y_0$) to determine the local skin friction. 

\begin{figure}
    \centering
    \includegraphics[width =\textwidth]{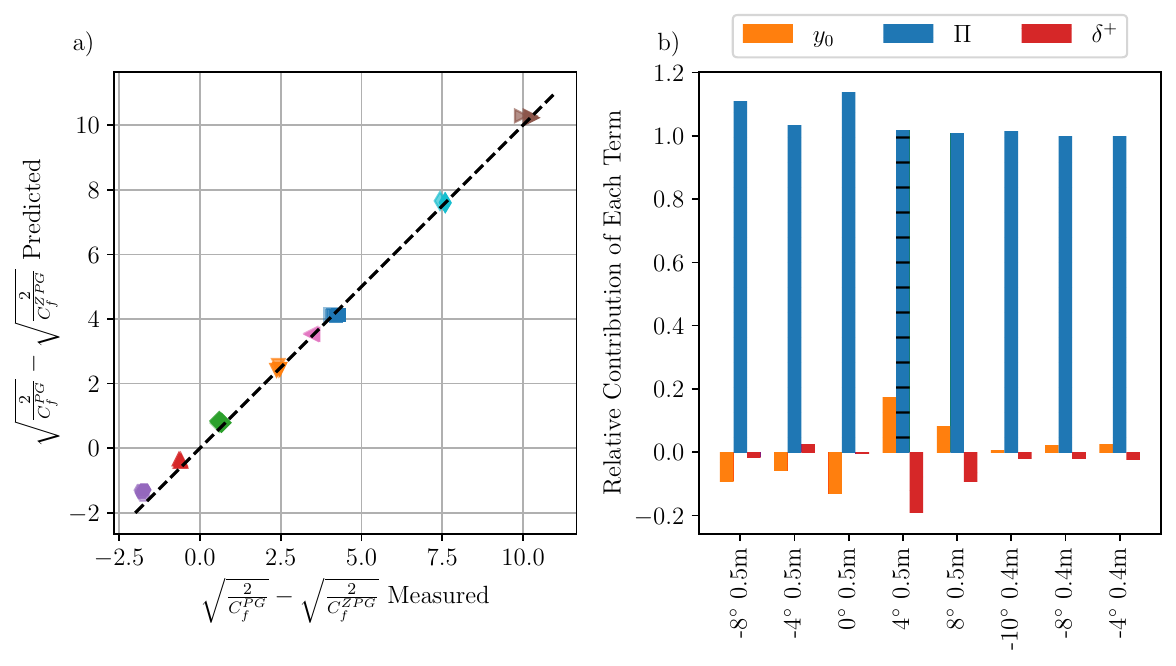}
    \caption{\textit{a)} Predicted difference in skin friction from equation \ref{eq:predict_cf_diff} against the measured skin friction difference from the drag balance for all Reynolds numbers and pressure gradients. The value of $\Pi$ is taken from fitting the velocity profile to equation \ref{eq:complete_velocity_profile_SW} and \ref{eq:complete_velocity_profile_RW}. The black dashed line is that of $y = x$, which would be true for a perfect prediction. \textit{b)} Relative contribution of each term in equation \ref{eq:predict_cf_diff} to the overall drag of the surface. Symbols and colours are as per figure \ref{fig:dCpdx}.}
    \label{fig:predicting_cf}
\end{figure}

The difference in skin friction from equation \ref{eq:predict_cf_diff} is plotted against the true difference in skin friction obtained from direct measurements in figure \ref{fig:predicting_cf}\textit{a} for all cases. It can be seen that there is an excellent agreement with all of the points lying along the diagonal line. Figure \ref{fig:predicting_cf}\textit{b} shows the relative contribution of each term of equation \ref{eq:predict_cf_diff} to the skin friction difference. As to be expected from the data presented in table \ref{tab:summary_data}, for a given freestream speed, the $\delta^+$ term is negligible and is seen to only have a marginal contribution to the overall skin friction. The contribution of the $y_0$ difference is also very small since any difference between them is negligible. The dominant contribution is from the $\Pi$ term as shown in figure \ref{fig:predicting_cf}\textit{b}. Equation \ref{eq:predict_cf_diff} therefore provides a solution to predict the skin friction increase due to an unknown pressure gradient history, if the wake strength ($\Pi$) is known.

We need to be able to predict $\Pi$ from a known flow history of pressure gradients in order to determine the skin friction. Following \cite{perry_2002}, the obvious parameter that can be used in this correlation is the pressure gradient parameter $\beta$. For our current problem, we cannot use $\beta$ since we only have data at a single streamwise location and $\beta$ at this location is zero (since local $dP/dx \approx 0$). Moreover, it will not be possible for us to evaluate $\overline{\beta}$ as proposed by \cite{vinuesa_2017} as the streamwise distribution of $\delta^*$ and $\tau_w$ is not available either (and it will not be for most studies as they are both outputs for a given $dP/dx$ history). Therefore, we introduce a new parameter, $\Delta \beta$ that can account for the pressure gradient history. This parameter is defined as, 

\begin{equation}
    \Delta \beta = \left(\frac{\delta^*}{\tau_w}\right)_{DS} \left[\frac{1}{x_{DS}-x_{US}}\int_{x_{US}}^{x_{DS}}\left(\frac{dP}{dx}\right) w(x) dx\right]~~{\rm where}~w(x) = \frac{x-x_{US}}{x_{DS}-x_{US}}
    \label{eq:delta_beta}
\end{equation}

where, $(\delta^*/\tau_w)_{DS}$ is the ratio of displacement thickness to the wall shear stress at the downstream measurement location, the integral term within $[\cdot]$ that includes $dP/dx$ distribution and a weighting function $w(x)$ is the weighted integral of streamwise pressure gradient history between the two streamwise locations. We hypothesise that local values of $\delta^*$ and $\tau_w$ at the measurement location already have history effects incorporated in them, and it may not be necessary to include them in the integral. Since the pressure gradient history closer to the measurement station has a greater influence compared to the upstream regions, some weighting should be applied to obtain a weighted pressure gradient history. For simplicity, a linear weighting as shown in equation \ref{eq:delta_beta} is applied. 

\begin{figure}
    \centering
    \includegraphics[width=0.5\linewidth]{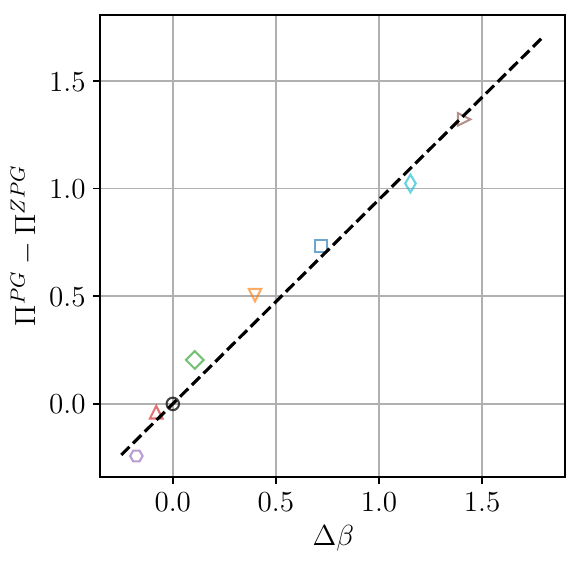}
    \caption{Difference between $\Pi^{PG}$ and $\Pi^{ZPG}$ for smooth wall as a function of $\Delta \beta$. Only the 20 m/s data is shown. Here the $\delta^*$ is calculated from the estimated profile with the near wall based on the Musker profile (\citealt{musker_1979}) and outer wake as given in equation \ref{eq:complete_velocity_profile_SW}. The black dashed line is the best fit to the data. Symbols and colours are as per figure \ref{fig:dCpdx}..  
     \protect\href{https://cocalc.com/share/public_paths/4319075c48d8c110898440464916f9827cd22b95/Figure\%2010/Figure\%2010.ipynb}{Notebook for Figure}}
        \label{fig:beta_pi}
\end{figure}

Using our experimental data for smooth walls, we develop a correlation between $\Delta \beta$ and $\Pi$ for the measurement location. Figure \ref{fig:beta_pi} shows $\Pi^{PG} - \Pi^{ZPG}$ as a function of $\Delta \beta$ just for the smooth wall case. Only the 20 m/s case is plotted here while the trend appears to hold for other freestream speeds as well. The blacked dash line gives the best fit linear relationship given by equation \ref{eq:best_fit_pi}. 

\begin{equation}
    \Pi^{PG} - \Pi^{ZPG} = 0.95\Delta \beta
    \label{eq:best_fit_pi}
\end{equation}

This suggests that once the weighted integral of the pressure gradient history (which can be provided as an input) is known, we can infer the local value of $\delta^*$, $\tau_w$, and $\Pi^{PG}$. This form is different to what previous works have found such as \cite{das_1987} who showed that local $\beta$ varies as $\Pi^2$. However, this would not appear to be case for the flows considered in the current study as local $\beta \approx 0$. The work of \cite{perry_2002} gave a different functional form using a theoretical relationship, $\beta = 0.5 + A\Pi^{4/3}$ based on the attached eddy hypothesis (where the $A$ is a constant derived from data). This relationship under predicts the value of $\Pi$ as it does not fully capture the non-equilibrium effects. In fact, \cite{perry_2002} included history effects, especially strong streamwise changes in $\beta$, through a gradient parameter $\zeta$ that captures $d\Pi/dx$. This gradient parameter together with the evolution equations (momentum integral) can be used to predict the streamwise evolution, which can be further calibrated using experimental data. In the current work, the history effects are captured with $\Delta\beta$ through integration of $dP/dx$ weighted by a function, $w(x)$, over a fixed streamwise distance where the pressure gradient effects are present. Changing this weighting function will result in an altered relationship between $\Delta \beta$ and $\Pi$. For example, an error function (rather than linear) that goes from $0$ at $x_{US}$ to $1$ at $x_{DS}$ (with 0.5 at the midpoint between $x_{US}$ and $x_{DS}$) lead to approximately half the slope (i.e. $\Pi^{PG} - \Pi^{ZPG} = 0.45 \Delta \beta$). However, it does not take away from the nature of the relationship between the two quantities. Comparing the current approach with that of \cite{perry_2002} is more involved, but the first steps towards this comparison is presented in Appendix \ref{appendix:a}. Further work is required to reconcile the similarities/differences between equation \ref{eq:best_fit_pi} (and various different weighting functions) and the work of \cite{perry_2002} which is beyond the scope of this study. 

\begin{figure}
    \centering
    \includegraphics[width=1\linewidth]{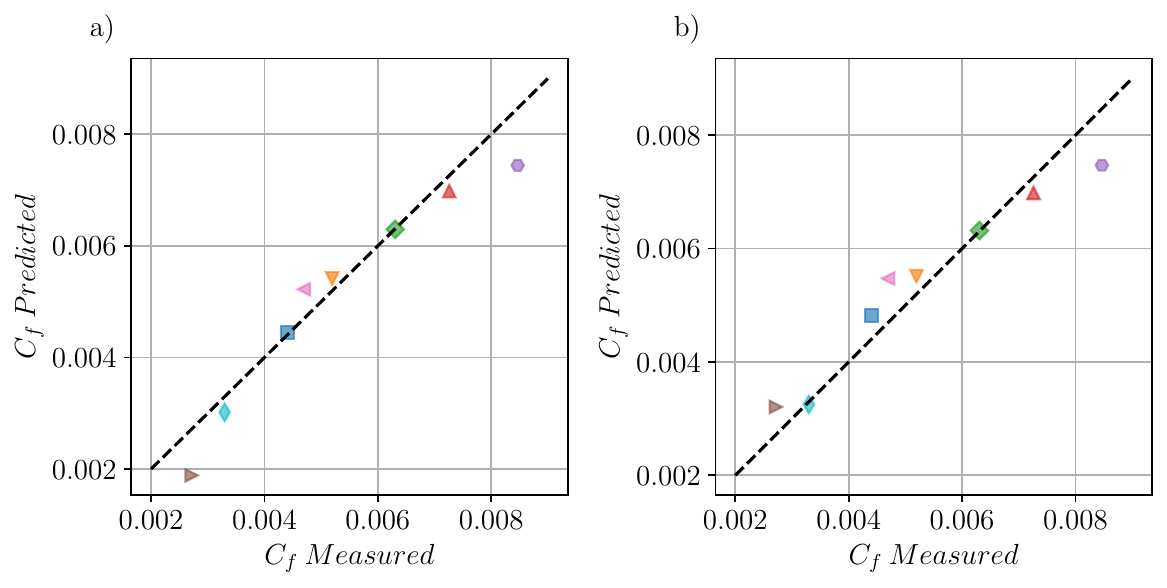}
    \caption{(\textit{a}) Predicted value of $C_f$ using minimisation function of equation \ref{eq:best_fit_pi} and the predicted fit of the velocity profile. $\delta^*$ here is provided as calculated from the hot wire velocity profile. This is compared to the measured value of $C_f$ with the black dashed line showing $y=x$, a perfect prediction. (\textit{b}) The predicted value of $C_f$ using the minimisation function of equation \ref{eq:best_fit_pi} and the predicted fit of the velocity profile. $\delta^*$ here is calculated using the velocity profile in \ref{eq:complete_velocity_profile_RW} where the value of $\Pi$ is implicitly included. The black dashed line shows $y=x$, which would be true for a perfect prediction. Symbols and colours are as per figure \ref{fig:dCpdx}.
    \protect\href{https://cocalc.com/share/public_paths/4319075c48d8c110898440464916f9827cd22b95/Figure\%2011/Figure\%2011.ipynb}{Notebook for Figure}}
    \label{fig:fmin_utau}
\end{figure}

We can test the predictive capability of the above relationship on flow over rough walls experiencing arbitrary pressure gradients as it is clear from figure \ref{fig:predicting_cf}\textit{b} that most of the skin friction comes from the changes in the wake parameter, especially if roughness length is not altered by the pressure gradient history. Given the weighted integral of the upstream pressure gradient history (as depicted in equation \ref{eq:delta_beta}), we solve equations \ref{eq:predict_cf_diff} and \ref{eq:best_fit_pi} simultaneously to obtain $C_f^{PG}$ and $\Pi^{PG}$ provided $C_f^{ZPG}$, $y_0$, and $\Pi^{ZPG}$ are known. Solving these equations also requires an input of $\delta^*$ or the velocity profile data at the location of prediction.~Figure \ref{fig:fmin_utau}\textit{a} shows the prediction of skin friction of rough walls experiencing different pressure gradients from this minimisation. The figure shows good agreement between prediction and data demonstrating the suitability of the derived correlations for both smooth and rough walls.~In fact, we can go a step further and include the calculation of $\delta^*$ as part of the minimisation process. In this case, the value of $\delta^*$ can be calculated from casting the mean velocity profile in equation \ref{eq:complete_velocity_profile_RW} in the appropriate form for displacement thickness, and therefore it depends only on input values for ZPG case (at matched $Re_\tau$). Consequently, $\delta^*$ can be implicitly determined as a part of the minimisation. Figure \ref{fig:fmin_utau}\textit{b} shows that the agreement between measured and predicted skin friction using this approach and it appears to be just as good as the former (where $\delta^*$ was given as an input). Overall, this shows the merit of the derived correlation-based method to predict skin friction of smooth and rough wall flows experiencing arbitrary pressure gradient histories. As such this method does have some limitations. Although the local values of $\beta$ over the pressure gradient history can attain large values (preliminary estimates show $\beta$ up to 5 in our flows), the range of integrated pressure gradient strengths is limited (i.e $0< \Delta \beta < 1.5$). We note that this integrated value is still higher than local $\beta$ achieved in some previous studies (\citealt{vishwanathan_2023_IJHF}). It is also unclear if the shape of the pressure gradient history is critical for this correlation. As such we are unable to validate this as most previous work do not have direct skin friction measurement or the scale separation required to use the proposed correlation. Exploring these effects should be considered in future works. 

\section{Conclusions}\label{sec:Conclusions}

Hot-wire and skin friction measurements have been presented for smooth and rough wall TBLs, showing the effect of non-equilibrium pressure gradients. Using a NACA0012 aerofoil, strong pressure gradient histories were created. The effects were investigated by measuring the velocity profiles one chord downstream of the trailing edge. Firstly, it was demonstrated that the pressure gradient is approximately zero one chord length upstream of the aerofoil. Consequently, the boundary layer one chord upstream remains invariant to the angle of attack. Therefore, any changes downstream of the wing are due to the different pressure gradient history. Several angles of attack of the wing produce different pressure gradient distributions along the streamwise direction. It was assumed that the pressure gradient type experienced closer to the measurement station was more dominant compared to the flow history further upstream. The direct skin friction measurements supported this conclusion. They showed that cases experiencing a favourable pressure gradient followed by an APG have lower skin friction compared to cases experiencing them in the reverse order. 

The velocity profiles taken one chord downstream of the aerofoil for both smooth and rough walls are at approximately matched $Re_\tau \approx 6800-8300$. There is a clear downward shift in the profiles for the rough wall compared to the smooth wall, presumably due to the additional drag. There is also a larger wake for APG immediately upstream, while an upstream FPG is found to suppress the wake. A key result was that the variation in the roughness length scale $y_0$ is not significant. The observed difference is assumed to be attributable to the fitting process and is thus considered invariant. Furthermore, it was seen that if $y_0$ is known from ZPG measurements, it is possible to predict the skin friction for the rough wall within 5\%. This error is comparable to the error observed in smooth wall log law fitting. 

The mean flow results enabled an investigation into how the effect of pressure gradients can be predicted in turbulent boundary layers. We examined modelling the difference in skin friction between a PG case and a ZPG case for a given surface. Inspired by the works of \cite{perry_2002}, \cite{castro_2007}, and \cite{vinuesa_2017} we developed a new correlation between the local skin friction and wake strength parameter, which in turn depends on a weighted-averaged pressure gradient parameter, $\Delta \beta$. The correlation was first trained on smooth wall data and then applied to flow over rough walls for predictions. We showed that this correlation can predict the skin friction of flows over smooth- and rough wall boundary layers with arbitrary pressure gradient history effects (given some input parameters from ZPG flows). Further work is necessary to understand the meaning of this new quantity and its ability to capture history effects. Work is also needed to capture the physical mechanisms (i.e. flow structure) that may be responsible for these observations.  

\backsection[Acknowledgements]{We gratefully acknowledge the help of Nick Agathangelou for taking the upstream smooth wall data as part of his undergraduate project.}

\backsection[Funding]{We gratefully acknowledge the financial support from EPSRC (Grant Ref no: EP/W026090/1) and European Office for Airforce Research and Development (Grant No: FA8655-22-1-7163, Programme Manager: Dr. Doug Smith). }

\backsection[Declaration of interests]{The authors report no conflict of interest.}

\backsection[Data availability statement]{All experimental data will be made available upon publication}

\backsection[Author ORCIDs]{T. Preskett: \href{https://orcid.org/0000-0001-9203-1266}{0000-0001-9203-1266}; M. Virgilio: \href{https://orcid.org/0009-0002-0603-4482
}{0009-0002-0603-4482}; \\ P. Jaiswal: \href{https://orcid.org/0000-0002-5240-9911}{0000-0002-5240-9911}; B. Ganapathisubramani: \href{https://orcid.org/0000-0001-9817-0486}{0000-0001-9817-0486}}

\backsection[Author contributions]{TP designed the setup, conducted experiments, and post-processed HWA and Balance data as well wrote the first draft. MV conducted experiments, post processed OFI data and edited drafts. PJ conducted experiments and edited drafts. BG provided the ideas for the experiment, acquired funding, edited drafts, and managed the project.}

\appendix
\section{Further details on the pressure coefficient and its gradient}
    \label{appendix:Cp_appendix}

\begin{figure}
    \centering
    \includegraphics[width = \textwidth]{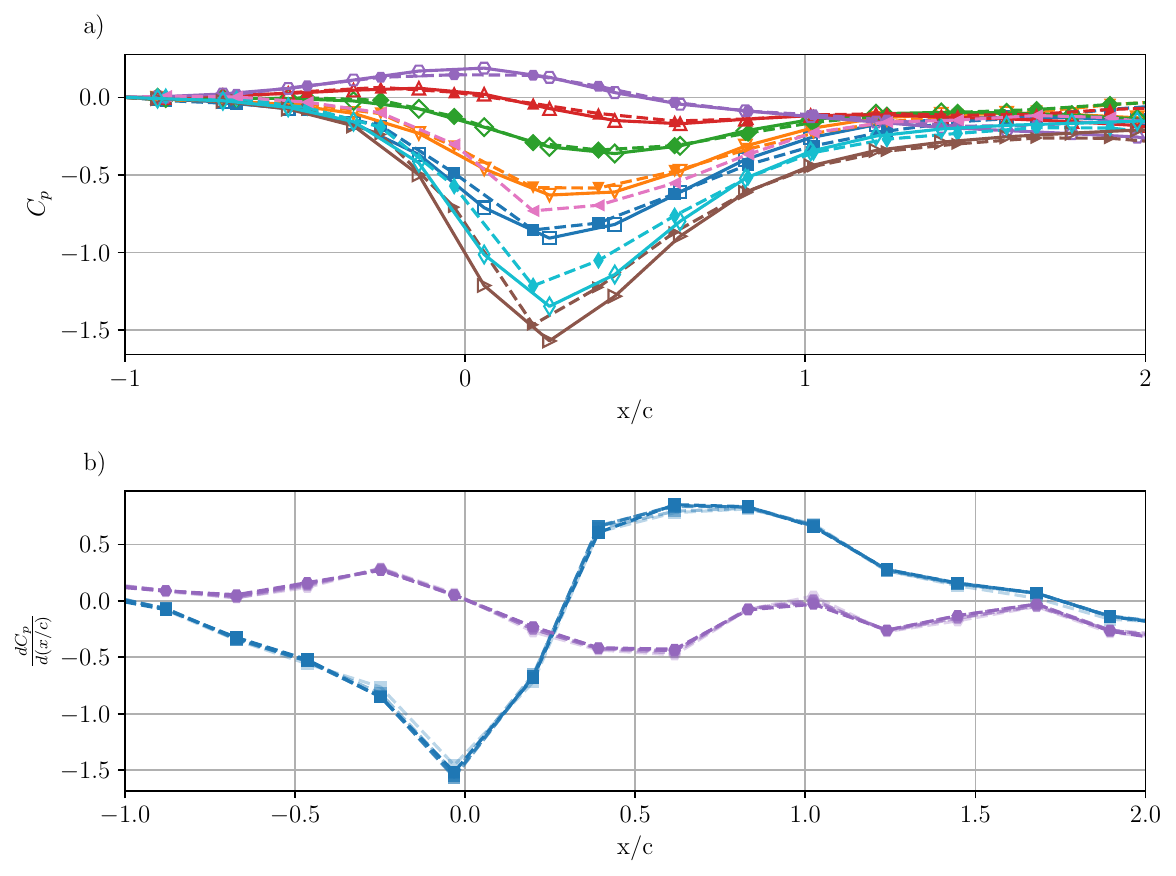}
    \caption{$C_p$ pressure distribution from one chord in front of the aerofoil to one chord behind the aerofoil for (\textit{a}) Mean $C_p$ pressure distribution history for the five cases at $h$ = 0.5 m for both smooth and rough walls. For the $h$ = 0.4 m cases, the three rough and two smooth wall cases are shown. Solid lines are used for smooth wall datasets, while dashed lines are used for rough wall data.  (\textit{b}) Rough wall $\frac{dC_p}{d(x/c)}$ for the $-8^\circ$ and $8^\circ$ cases with different Reynolds numbers. Symbols and colours are as per figure \ref{fig:dCpdx}.}
    \label{fig:Cp}
\end{figure}

The mean pressure distribution from all speeds is presented for smooth and rough walls at the five angles of attack for 0.5 m height in figure \ref{fig:Cp}\textit{a}. Also presented are the cases with a quarter chord height of 0.4 m. The pressure coefficient here is given by $C_p = (P-P_s) / (0.5 \rho U_\infty^2)$; here, $P_s$ is taken to be the static pressure at $x/c = -1$. It can be seen that both the smooth and rough wall datasets have very similar shapes. At 20 m/s the typical uncertainty in the $C_p$ value is $\pm 0.025$. Furthermore, the maximum pressure coefficient is located at the same streamwise location. There are slight differences in the pressure coefficient between the rough and smooth walls. The smooth wall for strong negative angles of attack has high magnitude peak values compared to the rough wall. The reason for this is likely due to the proximity of the taps to the roughness elements. This results in a lower pressure coefficient than for taps, which are further from roughness elements. 

Figure \ref{fig:Cp}\textit{b} shows the PG is shown for the $-8^\circ$ and $8^\circ$ cases for the rough wall. This shows an important result that the PG history is invariant to the Reynolds numbers, shown here for the most extreme cases. The small variations across the different speeds are due to minor differences in the boundary layer thicknesses, changing the tunnel's effective cross-section.

\section{First steps towards reconciling model presented in \cite{perry_2002} with the current work}
\label{appendix:a}

The work of \cite{perry_2002} presented a model which incorporates pressure gradient history effects into calculating $\Pi$. The key parameter for their work is $\zeta$, which is given by equation \ref{eq:zeta}. Using this parameter they define a relationship for $\beta$ which is shown in equation \ref{eq:beta_perry}. This gives $\beta = f(\Pi, d\Pi/dx)$ that for many flows presents a problem since knowing $d\Pi/dx$ is often not practical or possible to predict. It requires measurements at regular intervals with skin friction and boundary layer measurements at each station. 

\begin{equation}
    \zeta = \frac{U_{99}}{U_\tau} \delta \frac{d\Pi}{dx}
    \label{eq:zeta}
\end{equation}

\begin{equation}
  \beta =
    \begin{cases}
        -0.5 + 1.2\Pi^{\frac{4}{3}} + \zeta^2 (1.10/\Pi^2) & \text{if $\zeta \ge 0$}\\
       -0.5 + 1.2\Pi^{\frac{4}{3}} + \zeta (0.62 + 0.25 \Pi) & \text{if $\zeta < 0$}\\
    \end{cases}    
    \label{eq:beta_perry}
\end{equation}

Using the definition of $\beta$ it is possible to rewrite equation \ref{eq:beta_perry} in the form of equation \ref{eq:delta_beta} as given in equation \ref{eq:model_comparison}. The left hand side is the weighted integral of the pressure gradient and the right hand side comes from equation \ref{eq:beta_perry}.

\begin{equation}
\frac{1}{x_{DS}-x_{US}} \int^{x_{DS}}_{x_{US}} \frac{dP}{dx} w(x) dx = \frac{1}{x_{DS}-x_{US}} \int^{x_{DS}}_{x_{US}} \frac{\tau_w}{\delta^*}f\left(\Pi, \frac{d\Pi}{dx}\right) w(x) dx
\label{eq:model_comparison}
\end{equation}

where $f(\Pi, d\Pi/dx)$ is given by equation \ref{eq:beta_perry}.

\begin{figure}
    \centering
    \includegraphics[width=0.5\linewidth]{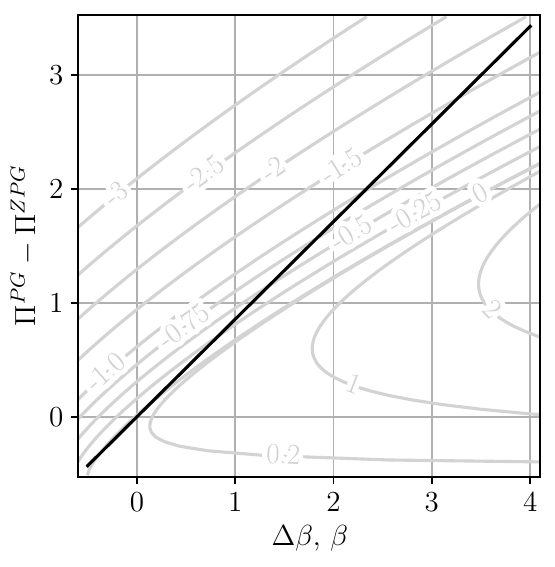}
    \caption{Comparison of variation $\Pi^{PG} - \Pi^{ZPG}$ with $\Delta \beta$ as given by equation \ref{eq:best_fit_pi} and variation in $\Pi^{PG} - \Pi^{ZPG}$ with $\beta$ from the model of \cite{perry_2002}. The black line is showing the fit of \ref{eq:best_fit_pi} and the labelled lines show variation of $\Pi_{PG} - \Pi_{ZPG}$ for different $\zeta$ values from equation \ref{eq:beta_perry}.}
    \label{fig:pi_model_comparison}
\end{figure}

Figure \ref{fig:pi_model_comparison} shows contours of $\Pi$ and $\beta$ from \cite{perry_2002} for various values of $\zeta$. A flow with a given pressure gradient history will trace out a path in this space. However, the $\Delta \beta$ term in the current study is a weighted integral of the pressure gradient term, and this will take on a value that can only be obtained through the integration of the function in this space. Figure \ref{fig:pi_model_comparison} also shows the curve that relates $\Delta \beta$ against $\Pi$ from equation \ref{eq:best_fit_pi}. It is clear that the relationship in equation \ref{eq:best_fit_pi} jumps from across different $\zeta$ curves for different pressure gradient histories. This can also be seen from equation \ref{eq:model_comparison} where $\Delta \beta$ is indeed an integral across different curves. The effect of this space is, in fact, captured through the weighting function, and at this stage, linear weighting appears to capture the trends reasonably well. However, further work is required to tune the weighting function for different pressure gradient histories and to reconcile the similarities/differences with \cite{perry_2002}.
 
\section{Comparison of current work with \cite{castro_2007}}
\label{appendix:b}

\cite{castro_2007} provided a three parameter model for skin friction $C_f = f(\frac{\theta}{y_0}, H, \Pi)$ over rough walls in ZPG conditions (where $H$ is the shape factor of the boundary layer and $\theta$ is the momentum thickness). This work showed that for a given value of $\Pi$ and H the variation in $C_f$ with $\theta/ y_0$ can be calculated using equation \ref{eq:cf_for_given_pi}. Alternately, it is also possible to obtain a variation of $C_f$ with the $H$ following for different values of $\Pi$ as shown below.

\begin{eqnarray}
    \sqrt{\frac{2}{C_f}} & = & -\frac{1}{\kappa} ln \left( \frac{1}{H} \sqrt{\frac{C_f}{2}} \right) + \frac{1}{\kappa} ln \left(\frac{\theta}{y_0}\right) + \frac{2 \Pi}{\kappa} - \frac{1}{\kappa} ln\left(\frac{1 +\Pi}{\kappa}\right) \label{eq:cf_for_given_pi} \\
   \sqrt{\frac{2}{C_f}} & = &  \left(\frac{H}{H-1}\right) \left[\frac{2.009 + 3.018 \Pi + 1.486 \Pi^2 }{\kappa (0.983 + \Pi)}\right] \label{eqn_H_Cf_formula}
\end{eqnarray}

\begin{figure}
    \centering
    \includegraphics[width=1\linewidth]{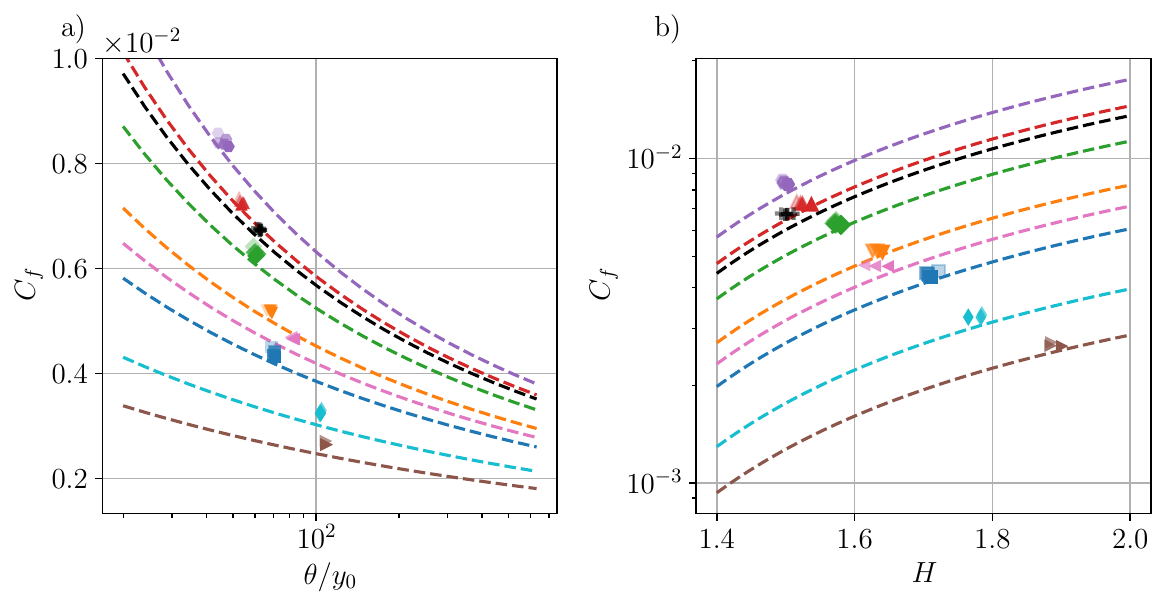}
    \caption{Three parameter model based on \cite{castro_2007} where it is assumed $C_f = f(\theta/y_0, H, \Pi)$. (\textit{a}) Variation in $C_f$ with $\theta/y_0$ curves show predicted $C_f$ variation for each cases $\Pi$ from equation \ref{eq:cf_for_given_pi}. (\textit{b}) Variation in $C_f$ with $H$ curves show predicted $C_f$ variation for each case $\Pi$ from equation \ref{eqn_H_Cf_formula}. Symbols and colours are as per figure \ref{fig:dCpdx}.}
    \label{fig:castro_figs}
\end{figure}

Note that the constants in equation \ref{eqn_H_Cf_formula} are taken directly from \cite{castro_2007} and they depend on the type of wake profile fitted to the velocity data. This is consistent with the use of \cite{lewkowicz_1982} polynomial wake profile, which is also used in the current study. 

This analysis can be extended for flows with pressure gradients. Figure \ref{fig:castro_figs}\textit{a} and \ref{fig:castro_figs}\textit{b} shows the predictions of $C_f$ as a function of $\theta/y_0$ as well as $C_f$ versus $H$ for the different rough wall flows examined in this study. Lines of constant $\Pi$ are shown in both figures and these lines are based on the value of $\Pi$ from the new correlation developed in section \ref{sec:Models} for different pressure gradient histories. 

Both figures show very good agreement between trendlines for different $\Pi$ and the skin friction measured. We also attempted to fit the wake function from \cite{cole_1956} wake function which does not seem to affect the fitted value of $\Pi$. However, in this case, the coefficients in equation \ref{eqn_H_Cf_formula} have to be altered. Regardless, this agreement shows that the method proposed in the current work is consistent with the results of \cite{castro_2007}, provided a suitable value of $\Pi$ as determined from $\Delta \beta$ is used.

 \bibliographystyle{jfm}
\bibliography{jfm}

\end{document}